\documentclass[twocolumn]{autart}    

\input{macros.tex}

\usepackage{cite}
\usepackage{amsmath,amssymb,amsfonts}
\usepackage{algorithm}
\let\classAND\AND
\let\AND\relax
\usepackage{algorithmic}

\let\AND\classAND
\AtBeginEnvironment{algorithmic}{\let\AND\algoAND}
\usepackage{graphicx}
\usepackage{textcomp}
\usepackage{bm}
\usepackage[inline]{enumitem}
\usepackage{todonotes}
\usepackage{wrapfig}
\usepackage{thm-restate}
\usepackage{booktabs}
\usepackage{caption}
\usepackage{array}
\usepackage{multicol}
\usepackage{tcolorbox}
\usepackage{multirow}
\usepackage{makecell}
\usepackage{pifont}
\usepackage{fontawesome5}
\usepackage{xcolor}

\usepackage{pgf}
\usepackage{tikz}
\usetikzlibrary{arrows,shapes,fit,tikzmark,snakes,automata,backgrounds,petri,calc,hobby,positioning,fadings,through,arrows.meta,decorations.pathmorphing}
\usepackage{etoolbox}
\usepackage{adjustbox}
\usetikzlibrary{patterns}

\input{tikz.tex}
\usepackage{graphicx}          

\begin{document}

\begin{frontmatter}

\title{Reset Controller Synthesis for Delay Hybrid Systems using Reach-avoid Analysis} 


\author[ISCAS]{Han Su}\ead{suhan@ios.ac.cn},                      
\author[ISCAS]{Jiyu Zhu}\ead{zhujy@ios.ac.cn}, 
\author[ZGCL]{Shenghua Feng}\ead{fengsh@zgclab.edu.cn},            
\author[BICE]{Yunjun Bai}\ead{baiyj@ios.ac.cn},
\author[BICE]{Bin Gu}\ead{gubinbj@sina.com},  
\author[CIGIT]{Jiang Liu}\ead{liujiang@cigit.ac.cn},  
\author[CAST]{Mengfei Yang}\ead{yangmf@bice.org.cn},  
\author[PKU,ISCAS]{Naijun Zhan}\ead{znj@ios.ac.cn}

\address[ISCAS]{Institute of Software, University of Chinese Academy of Sciences, Beijing, China}  
\address[ZGCL]{Zhongguancun Laboratory, Beijing,China}                                      
\address[BICE]{Beijing Institute of Control Engineering, Beijing, China}             
\address[CIGIT]{Chongqing Institute of Green and Intelligent Technology, Chinese Academy of Sciences, Chongqing, China}        
\address[CAST]{China Academy of Space Technology, Beijing, China}
\address[PKU]{School of Computer Science, Peking University, Beijing, China}
          

\begin{abstract}
A reset controller is pivotal in the design of hybrid systems. It restricts the initial set and redefines the reset map associated with discrete transitions, ensuring system objective is met. Reset controller synthesis, alongside feedback controller synthesis and switching logic controller synthesis, provides a correct-by-construction approach to designing hybrid systems. However, the presence of time-delay poses challenges in hybrid systems, potentially compromising control performance and rendering verification certificates obtained by abstracting away time-delay invalid in practice. This paper addresses this issue by proposing a approach that incorporates time-delay considerations. We introduce a method that reduces the synthesis of reset controllers to the generation of reach-avoid sets for the hybrid system at hand, which can be efficiently solved using standard convex optimization solvers.
\end{abstract}

\end{frontmatter}

\section{Introduction}

Hybrid systems (HSs) offer a robust mathematical framework for modeling cyber-physical systems (CPS), seamlessly integrating continuous physical dynamics with discrete switching behaviors. The correct design of reliable HSs is a critical research area, particularly in safety-critical domains like healthcare and medicine \cite{dey2018medical}, autonomous vehicles \cite{chattopadhyay2017security}, and automated factories \cite{zhao2012hybrid}. However, as CPSs grow increasingly complex, the inevitability of time delays poses additional challenges to the correct design of hybrid systems.

Typically, HSs that account for delays are termed Delay Hybrid Systems (dHS). A dHS comprises two types of delay: one manifests in the continuous evolution of systems, where the evolution depends not only on the current state but also on the historical state. This delay is often modeled using delay differential equations (DDEs). The other type of delay arises during discrete transitions between different control modes of the dHS.

Controller synthesis provides a correct-by-construction manner to construct an operational behavior model ensuring that a given dHS adheres to specified properties like safety and reach-avoid. Three common control mechanisms for dHS include \emph{feedback controllers}, \emph{switching logic controllers}, and \emph{reset controllers}. Feedback and switching logic controller synthesis have been extensively studied over decades, as evidenced by works such as \cite{sanfelice2021hybrid, branicky1994stability, zhao2019optimal, tomlin2000game, ames2016control, girard2012controller, tabuada2009verification} for feedback controllers and \cite{tabuada2009verification, belta2017formal, reissig2016feedback, nilsson2017augmented, hsu2018multi, zhao2013synthesizing, taly2011synthesizing} for switching logic controllers. However, reset controller synthesis is surprisingly overlooked in the literature, and even worse, most of existing work assume delay-free dynamics. 

Reset controller synthesis, through the redesign of the reset map and initial conditions, provides an effective solution for ensuring the correctness of dHS designs. In certain situations, reset controllers can offer greater efficacy compared to other control mechanisms. For instance, during a transition from one mode to another in a dHS, the historical sensor data is utilized to initialize the delay feedback controller in the new mode. However, this historical data might lead to behaviors that deviate from the control objectives in the new mode. In such cases, if we can use a new collection of data, which can be seen as the initial condition of the DDE in the new mode, to replace the history data that the controller in the new mode will use when a switch happens, or saying, if we can modify an appropriate reset map after discrete transition, we may achieve the control goal in a more simple manner.

In this paper, we present our investigation into the synthesis of reset controllers for dHS. The reset map associated with a discrete transition is typically a set-valued function that defines the relationship between continuous evolution in the post-mode and the previous mode. Our goal is to synthesize a reset map ensuring both safety and liveness conditions for dHS. Specifically, we aim to find a reset map ensuring that all dHS executions reach a target set $\Target$ while staying within a safe set $\Safe$. To efficiently address this, we introduce a novel reach-avoid analysis method for DDEs, leveraging a reach-avoid barrier functional (RABF). We demonstrate that RABF can be synthesized by solving a reduced \emph{semidefinite programming} (SDP) problem \cite{wolkowicz2012handbook}. The 0-sublevel set of RABF offers an inner approximation of the reach-avoid set. Subsequently, we propose a two-step approach for reset controller synthesis in dHS. 
\begin{enumerate*}[label=(\roman*)]
    \item We decouple continuous and discrete behaviors in dHS using the reach-avoid analysis method, transforming dHS into a discrete directed graph (DDG) by eliminating continuous dynamics.  
    \item We identify and block edges potentially leading to \emph{``non-target sink''} or \emph{``infinite loop''} scenarios based on the resulting DDG, where dHS fails to reach the target set. Based on the pruned DDG, we synthesize a reset controller to ensure safety and liveness for dHS. 
\end{enumerate*}
Experimental results on literature examples validate the effectiveness of our approach.


In summary, our main contributions are as follows.
\begin{itemize}
   \item We propose a novel method for reach-avoid analysis of DDEs, achieving better performance than existing methods.
   \item We introduce an efficient approach to synthesizing reset controllers for dHS by reducing the problem to reach-avoid analysis for continuous dynamics and simple loop analysis using depth-first search for discrete dynamics. 
   \item We provide a prototypical implementation of our approach and apply it to several case studies, demonstrating its effectiveness and efficiency.
\end{itemize}

\paragraph*{\textbf{Organization.}} In the following, Sect.\nobreakspace \ref{sec:pre} provides a recap of important preliminary definitions and formally defines the problem of interest. Sect.\nobreakspace \ref{sec:continuous} presents our reach-avoid analysis method by introducing RABF, Sect.\nobreakspace \ref{sec:consDG} proposes our reset controller synthesis method by constructing the DDG and pruning the edges. In Sect.\nobreakspace \ref{sec:exper}, we demonstrate the effectiveness of our method through several examples. Finally, we conclude the paper in Sect.\nobreakspace \ref{sec:conclu}.

\section{Preliminaries and Problem Formulation}\label{sec:pre}

    \paragraph*{\textbf{Notations}} Let $\Real$, $\PosReal$, and $\Realn$ denote the set of real numbers, non-negative real numbers, and $n$-dimensional real numbers, respectively. $\ContiFun{[-\tau,0]}{\Realn}$ represents the Banach space of continuous functions that map interval $[-\tau,0]$ to $\Realn$, equipped with the norm $\ContiNorm{\x_t} = \sup_{\theta\in[-\tau,0]}||\x_t(\theta)||$, where $||\cdot||$ represents the Euclidean norm. Given set $A$, let $\overline{A}$, $\partial A$, and $\PowSet{A}$ represent its closure, boundary, and power set, respectively. $\Poly{\x}$ denotes the polynomial ring in $\x$ over the field $\Real$. $\Poly{\x}^n$ denotes the set of $n$-dimensional vectors, where each element is a polynomial in $\x$ over the field $\Real$. All vectors in this article are considered as column vectors by default, and $A^T$ denotes the transpose of  $A$. Given any vector $\textbf{u}$, $|\textbf{u}|$ denotes its coordinate-wise absolute value. 



    \subsection{Delay Differential Equation}\label{subsec:dde}
        We consider a class of dynamic systems featuring differential dynamics governed by DDEs of the form~\cite{fridman2014introduction}
        \begin{align}\label{eq:dde}
                \dot \x(t) = \f(\x(t),\x(t-\tau)),\quad 0<\tau<\infty
        \end{align} 
        where $\f$ is a continuous differentiable function. Given initial condition $\pphi(\cdot)\in\ContiFun{[-\tau,0]}{\Realn}$, there exists an unique solution (or trajectory) 
        $\x^{\pphi}(\cdot):[-\tau,+\infty]\to \Realn$, such that $\dot \x^{\pphi}(t) = \f(\x^{\pphi}(t),\x^{\pphi}(t-\tau))$ for $t\ge 0$, and $\x(t)=\pphi(t)$ for $t\in[-\tau,0]$.

        Unlike ordinary differential equations (ODEs), delay differential equations are essentially functional equations: the evolution of DDEs depends not only on the current state but also on the historical state. To concisely represent the dynamics of the system's history, we utilize the function $\x_t(\cdot):[-\tau,0]\to\Realn$, which denotes the values of $\x$ from time $t$ to $t-\tau$. Specifically, $\x_t(\theta)=\x(t+\theta)$ for $\theta\in[-\tau,0]$.

    \subsection{Delay Hybrid Automata}\label{subsec:DHA}
        Delay hybrid automata (dHA) \cite{bai2021switching} extends the notion of classical hybrid automata (HA) by incorporating delays, providing an approximate  mathematical model for characterizing the behavior of dHS. In this paper, we employ dHA as the foundational model.

        \begin{defn}[Delay Hybrid Automata \cite{bai2021switching}]\label{df:edha}
            A Delay Hybrid Automaton, denoted as $\bm{\mathcal{H}}=(\DisState, \ContiState, \Dom, \ContiDyn,\allowbreak \Init, \DisJump, \Guard, \STime, \Reset)$, consists of the following components:
            \begin{itemize}
                \item $\DisState = \{q_1, q_2, \ldots, q_m\}$, a finite set of modes.
                \item $\ContiState$, a set of continuous state variables. A continuous state $\x$ is a valuation of all variables in $\ContiState$.
                \item $\Dom: \DisState \to \PowSet{\Realn}$, a function assigning to each mode an invariant domain within which the system operates.
                \item $\ContiDyn = \{\f_1, \f_2, \ldots, \f_m\}$, a set of vector fields, each corresponding to a mode. For each mode $q_i$, the continuous dynamics is governed by the DDE:
                \[
                \dot\x(t) = \f_i(\x(t), \x(t-\tau_i));
                \]
                \item $\Init: \DisState \to \PowSet{\ContiFun{[t_1, t_2]}{\Realn}}$, where $t_1 \leq t_2$, maps each mode to a set of initial conditions.
                \item $\DisJump \subseteq \DisState \times \DisState$, a set of discrete transitions between modes\footnote{For a transition $e=(q_1,q_2)$ in $\edHAabb$, we call $q_1$ the pre-mode of $e$ while $q_2$ the post-mode of $e$.}.
                \item $\Guard: \DisJump \to \PowSet{\Realn}$, a function specifying the guard conditions that enable transitions between modes.
                \item $\STime: \DisJump \to \PosReal$, a function indicating the time duration associated with each discrete transition.
                \item $\Reset: (\DisJump \times \ContiState) \to \PowSet{\ContiFun{[t_1, t_2]}{\Realn}}$, a function defining the reset map for each transition $e = (q_1, q_2)$, mapping a continuous state $x \in \Realn$ from the pre-mode $q_1$ to a set of initial conditions in the post-mode $q_2$.
            \end{itemize}
            \end{defn}

        The \emph{instantaneous state} of a dHA at any given time instant is a tuple $(q,\x)$, specifying a mode $q$ and a \emph{continuous state} $\x$. The state can change in two ways:
        \begin{enumerate*}[label=(\roman*)]
            \item by a transition that changes the entire state according to the discrete relation and reset map, and
            \item by elapse of time that changes only the continuous state according to the DDE defined by the vector field of the current mode.
        \end{enumerate*}  
        An execution of a dHA is formally defined as follows:

        \begin{defn}[Hybrid Execution]\label{def:HEX}
            For a dHA $\edHAsyb$ with initial state $(q_0,\pphi_0(\cdot))$, a hybrid execution $\pi$ consists of a sequence of triples $\langle t_i,q_i,\x^{\pphi_i}(t_i)\rangle$, where $i\in\Nat$ and $q_i\in\DisState$. Each transition $\langle t_i,q_i,\x^{\pphi_i}(t_i) \rangle \mapsto \langle t_{i+1},q_{i+1},\x^{\pphi_{i+1}}(t_{i+1}) \rangle $ is either
            \begin{itemize}
                \item discrete transition: $e=(q_i,\allowbreak q_{i+1})\allowbreak \in\DisJump$, $t_{i+1} = t_i + \STime(e)$, $\x^{\pphi_i}(t_i)\in\Guard(e)$,  $\pphi_{i+1}\in\Reset(e,\x^{\pphi_i}(t_i))$, or
                \item continuous evolution: $q_i=q_{i+1}$, $\pphi_{i}=\pphi_{i+1}$, $t_i < t_{i+1}$, with $\x^{\pphi_i}(\cdot)$ evolves according to DDE defined by $\f_i$ over interval $[t_i,t_{i+1}]$, and $\x^{\pphi_i}(t)\in \Dom(q_i)$ for $t\in[t_i,t_{i+1}]$. 
            \end{itemize}
        \end{defn}

        \begin{rem}
            A discrete transition may not happen immediately upon satisfying the guard condition, but it must occur instantly when the invariant domain is breached. This requirement mirrors practical systems, where the continuous state in each mode is confined to a specific domain (i.e., invariant domain). If the continuous state exceeds this domain, the system will switch to another mode, otherwise the system will become blocked.
        \end{rem}

        \begin{rem}
        The reset map is not a physical process but rather a logical construct within the model. It specifies an initial condition that may not occur in reality, serving instead as a virtual starting point. Such initial condition can be utilized to initiate the delay feedback controller in the subsequent mode or to guide the design of practical hybrid systems
            
        \end{rem}

        Serving as a mathematical model, dHA can guide the practical construction of real systems, providing formal guarantees for various control objectives. Here, we present an example to demonstrate the different components of dHA.

        \begin{figure*}[t]
            \centering
            \begin{adjustbox}{max width=1\linewidth}
            \scalebox{1}{
                \begin{tikzpicture}[font=\tiny]
                    \begin{groupplot}[ Axis Set, width = 4cm , height = 4cm , group style={group size = 3 by 1, horizontal sep = 4.5cm, vertical sep = 0.15cm}]
                        \nextgroupplot[xmin = -1.3, xmax=1.3, ymin = -1.3, ymax=1.3,xlabel = $x_1$, ylabel=$x_2$,ytick={1},xtick=\empty]
                        \addplot [name path = INV, domain=0:360, samples=100, thick, color = myblue] ({cos(x)},{sin(x)}); 
                        \addplot [name path = G_Q3,domain=0:360, samples=100, thick, color = mygreen] ({-0.3+sqrt{0.1}*cos(x)},{-0.3+sqrt(0.1)*sin(x)});
                        \addplot [name path = G_Q2,domain=0:360, samples=100, thick, color = myorange] ({sqrt{0.1}*cos(x)},{sqrt(0.1)*sin(x)});
                        \node at(axis cs:0.5,0.5) {\textcolor{myblue}{$\Dom(q_1)$}};
                        \node at(axis cs:-0.3,-0.3) {\textcolor{mygreen}{$\Guard(e_2\!)$}};
                        \node at(axis cs:0,0) {\textcolor{myorange}{$\Guard(e_1\!)$}};
                     
                        \nextgroupplot[xmin = -1.3, xmax=1.3, ymin = -1.3, ymax=1.3,xlabel = $x_1$, ylabel=$x_2$,ytick={1},xtick=\empty]
                        \addplot [name path = INV, domain=0:360, samples=100, thick, color = myorange] ({cos(x)},{sin(x)}); 
                        \node at(axis cs:0.5,0.5) {\textcolor{myorange}{$\Dom(q_2)$}};
    
                        \nextgroupplot[xmin = -1.3, xmax=1.3, ymin = -1.3, ymax=1.3,xlabel = $x_1$, ylabel=$x_2$,ytick={1},xtick=\empty]
                        \addplot [name path = INV, domain=0:360, samples=100, thick, color = mygreen] ({cos(x)},{sin(x)}); 
                        \addplot [name path = G_Q1, domain=0:360, samples=100, thick, color = myblue] ({0.5+sqrt(0.1)*cos(x)},{0.5+sqrt(0.1)*sin(x)}); 
                        \node at(axis cs:-0.3,0.5) {\textcolor{mygreen}{$\Dom(q_2)$}};
                        \node at(axis cs:0.5,0.5) {\textcolor{myblue}{$\Guard(e_3\!)$}};
                    \end{groupplot} 

                    \begin{scope}
                        \draw[thick,rounded corners] (-0.6,2.6) |- (3.2,-1.7) -- (3.2,2.6) -- cycle; 
                        \node at(-0.4,2.4) {\textcolor{myblue}{\scriptsize{\bm{$q_1$}}}};
                        \node at(1.3,-0.7) {$
                            \begin{aligned}
                                &\dot x_1(t)\!=\!0.5 x_2(t)\!+\! 0.5 x_2(t\!-\!0.1),\\
                                &\dot x_2(t)\!=\!-\!0.5 x_1(t) \!-\! 0.5 x_1(t\!-\!0.1)\\
                                &\qquad\quad\qquad\qquad\qquad -\! 1.5 x_2(t),
                            \end{aligned}$};
                        \node at(1.3,-1.37) {$\Init(q_1)=\ContiFun{[-0.1,0]}{\Dom(q_1)}$};
                    \end{scope}
                    
                    \begin{scope}[xshift=6.85cm]
                        \draw[thick,rounded corners] (-0.65,2.6) |- (3.25,-1.7) -- (3.25,2.6) -- cycle; 
                        \node at(-0.4,2.4) {\textcolor{myorange}{\scriptsize{\bm{$q_2$}}}};
                        \node at(1.3,-0.7) {$
                            \begin{aligned}
                                &\dot x_1(t)\!=\!0.88 x_2(t)\!+\! 0.12 x_2(t\!-\!1),\\
                                &\dot x_2(t)\!=\!-\!0.88 x_1\!(t)\!-\!0.12 x_1\!(t\!-\!1)\\
                                &\qquad\quad\qquad\qquad\qquad -\!1.5 x_2(t),  
                            \end{aligned}$};
                        \node at(1.3,-1.37) {$\Init(q_2)=\emptyset$};
                    \end{scope}

                    \begin{scope}[xshift=13.7cm]
                        \draw[thick,rounded corners] (-0.6,2.6) |- (3.2,-1.7) -- (3.2,2.6) -- cycle; 
                        \node at(-0.4,2.4) {\textcolor{mygreen}{\scriptsize{\bm{$q_3$}}}};
                        \node at(1.3,-0.7) {$
                            \begin{aligned}
                                &\dot x_1(t)\!=\!0.6 x_2(t)\!+\!0.4 x_2(t\!-\!0.2),\\
                                &\dot x_2(t)\!=\!-\!0.6 x_1(t)\!-\!0.4 x_1(t\!-\!0.2)\\
                                & \qquad\quad\qquad\qquad\qquad -\! 1.5 x_2(t),
                            \end{aligned}$};
                        \node at(1.3,-1.37) {$\Init(q_1)=\emptyset$};
                    \end{scope}

                    \draw[-stealth, thick] (3.2,1.5) -- (6.2,1.5) 
                        node [midway, yshift=.0cm, inner sep=1pt] {
                            $\begin{array}{c}
                                e_1=(q_1,q_2),\\
                                \Guard(e_1),\STime(e_1)=1,\\
                                \Reset(e_1,(x_1,x_2))=\\
                                \qquad\ContiFun{[-1,0]}{\Dom(q_2)}
                            \end{array}$};
                    \draw[-stealth, thick] (1.27,2.6) -- (1.27,3) -- (15.3,3) 
                        node [midway, yshift=.2cm, inner sep = 1pt] {
                            $e_2=(q_1,q_3), \Guard(e_2),\STime(e_2)=1, \Reset(e_2,(x_1,x_2))=\ContiFun{[-0.2,0]}{\Dom(q_3)}$
                        } -- (15.3,2.6);
                    \draw[-stealth, thick] (15.3,-1.7) -- (15.3,-2.1) -- (1.27,-2.1) 
                        node [midway, yshift=.2cm, inner sep = 1pt] {
                            $e_3=(q_3,q_1), \Guard(e_3),\STime(e_3)=1, \Reset(e_3,(x_1,x_2))=\ContiFun{[-0.1,0]}{\Dom(q_1)}$
                        } -- (1.27,-1.7);
                \end{tikzpicture}
                    }
                \end{adjustbox}
            \caption{The dHA for hybrid delay damped oscillator} 
            \label{fig:hldo}
        \end{figure*}
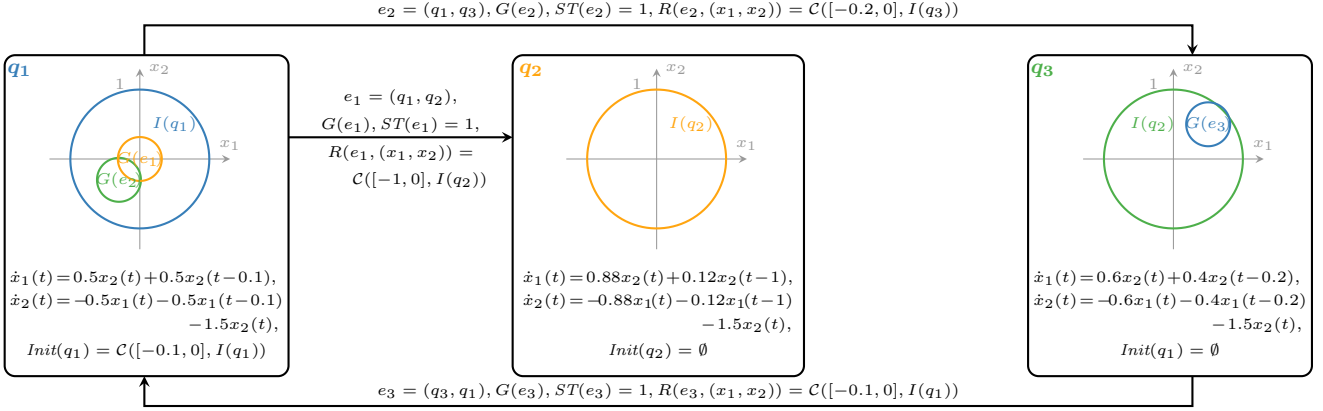

        \begin{exmp}\label{exp:running}
            Consider a hybrid delay damped oscillator capable of switching between different damping ratios. It operates in three modes $q_1,q_2,q_3$, each defined by distinct DDEs.
            \begin{align*}
                q_1: &\left\{~\begin{aligned}
                    &\dot x_1(t) =  0.5 x_2(t) + 0.5 x_2(t-0.1),\\
                    &\dot x_2(t) = -0.5 x_1(t) - 0.5 x_1(t-0.1) - 1.5 x_2(t), 
                \end{aligned}\right.\\
                q_2: &\left\{~\begin{aligned}
                    &\dot x_1(t) =  0.88 x_2(t) + 0.12 x_2(t-1),\\
                    &\dot x_2(t) = -0.88 x_1(t) - 0.12 x_1(t-0.1) - 1.5 x_2(t), 
                \end{aligned}\right.\\
                q_3: &\left\{~\begin{aligned}
                    &\dot x_1(t) =  0.6 x_2(t) + 0.4 x_2(t-0.2),\\
                    &\dot x_2(t) = -0.6 x_1(t) - 0.4 x_1(t-0.2) - 1.5 x_2(t),
                \end{aligned}\right.\\
            \end{align*} 
            where $x_1$ and $x_2$ represent the horizontal and vertical positions of the oscillator, respectively.

            The reset map for each discrete transition is universally valid. Specifically, for a discrete transition $e_1=(q_1,q_2)$ and continuous state $(x_1,x_2) \in \Guard(e_1)$, the function mapping $e_1$ and $(x_1,x_2)$ to any $\phi_0 \in \ContiFun{[-\tau_2,0]}{\Dom(q_2)}$ is a correct reset map. $\phi_0$ then serves as the initial condition for the dynamics in mode $q_2$. The dHA of this system is graphically illustrated in Fig.\ref{fig:hldo}, highlighting each component.
        \end{exmp}

        Given the established definitions and notations, we define the reachable set $\R$ of dHA as follows:
        \begin{defn}[Reachable Set]
            A state $(q,\x)$ of $\edHAabb$ is called reachable if there exists an execution ending at $(q,\x)$. The reachable set $\R$ comprises all reachable states, formally:
            \begin{align*}
                \R = \left\{\! (q,\x)\! \left| 
                    \begin{aligned}
                         \exists \pi= \langle t_0,q_0,\x^{\pphi_0}(t_0)\rangle,\cdots,\langle t_i,q_i,\x^{\pphi_i}(t_i)\rangle,\\
                        \pphi_0\in\Init(q_0), \text{and } (q_i,\x^{\pphi_i}(t_i))=(q,\x)
                    \end{aligned}
                    \right.\right\}
            \end{align*}
        \end{defn}

    \subsection{Problem Formulation}\label{sec:prob}
        Given a dHA $\edHAsyb$ as defined in Definition\nobreakspace\ref {df:edha} and a set of states $\Safe$. 
        A dHA $\edHAsyb$ is \emph{safe} with respect to $\Safe$ if every reachable state $(q,\x)$ is contained within $\Safe$. The problem we aim to address is formulated as follows:
        \begin{tcolorbox}[boxrule=.5pt,colback=white,colframe=black!75]
            \textbf{Reset Controller Synthesis.} Given a dHA $\edHAsyb$ as specified in Definition\nobreakspace\ref {df:edha}, a bounded safe set $\Safe\subseteq \DisState\times\Realn$, and a target set $\Target\subseteq\Safe$, can we identify a revised initialization function $\Init^r$ and reset function $\Reset^r$ such that all executions of the modified dHA $\edHAsyb^r=(\DisState,\ContiState,\Dom,\ContiDyn,\Init^r,\DisJump,\Guard,\STime,\Reset^r)$ remain within $\Safe$ until reaching $\Target$?
        \end{tcolorbox}

            

\section{Reach-avoid Analysis for DDE}\label{sec:continuous}
    To synthesize the reset controller for dHA according to the reach-avoid specification, it's crucial to initially analyze the reach-avoid problem within each mode, which aligns with solving the reach-avoid problem of DDEs.

    In the realm of DDEs, a reach-avoid set $\RA$ comprises initial conditions from which the trajectories must reach the target set $\Target$ within a finite time while remaining within the safe set $\Safe$ until they hit the target. Formally, the reach-avoid set is defined as:
    \begin{align*}
      \RA\DefSym\!\left\{\!\pphi\in \ContiFun{[-\tau,0]}{\Safe}\left|
        \begin{aligned} 
            &\exists\,t'\in\PosReal,~ \x^{\pphi}(t')\in\Target\wedge\\
            &\forall\,t\in[-\tau,t'),~\x^{\pphi}(t)\in\Safe~
        \end{aligned}\right.\!\right\}.
    \end{align*}
    In this section, for simplicity, $\Safe$ and $\Target$ refer to subsets of $\Realn$, rather than state sets of dHA. Consequently, the reach-avoid problem for DDE can be defined as below:
    \begin{tcolorbox}[boxrule=.5pt,colback=white,colframe=black!75]
        \textbf{Inner-Approximate Reach-Avoid Set.}Given $\Safe\subseteq \Realn$ a bounded safe set and $\Target\subseteq\Safe$ a target set, we aim to inner-approximate the reach-avoid set $\RA$. 
    \end{tcolorbox}



    \subsection{Reach-Avoid Barrier Functional}\label{subsec:ra}
        To more precisely inner-approximate $\RA$, inspired by \cite{kiss2022control} and \cite{xue2021reach}, we propose the notion of \emph{reach-avoid barrier functional} (RABF), whose 0-sublevel set provides an inner-approximation of $\RA$. Formally, 
        \begin{defn}[Reach-Avoid Barrier Functional]\label{df:rabfal}
            Given a DDE in the form of \eqref{eq:dde}, and safe set $\Safe$ and target set $\Target$ defined by 
            \begin{align*}
                \Safe \DefSym \{\x\in\Realn \mid \fsafe(\x) \le 0 \},  \,  \Target \DefSym \{\x\in\Realn \mid \ftarget(\x) \le 0\}\, ,
            \end{align*}
            we call the continuous differential functional $H(\cdot):\ContiFun{[-\tau,0]}{\Realn}\to\Real$ a \emph{reach-avoid barrier functional} if there exists a continuously differentiable function $w(\cdot):\Realn\to\Real$ such that the following conditions are satisfied:
            \begin{subequations}\label{condi:main-rabf}
                \begin{align}
                    &\frac{d H(\x_t)}{d t}  \le 0, ~ \forall\,\x_t\in\ContiFun{[-\tau,0]}{\overline{\Safe\!\setminus\!\Target}}, \label{cond:deriv} \\
                    &H(\x_t)                \ge 0,  ~ \forall\,\x_t\in\ContiFun{[-\tau,0]}{\overline{\Safe\!\setminus\!\Target}},\x_t(0)\in\partial\Safe,\label{cond:boundary}\\
                    &H(\x_t) \ge \frac{d w(\x_t(0))}{dt} ,~\forall\,\x_t\in\ContiFun{[-\tau,0]}{\overline{\Safe\!\setminus\!\Target}} .\label{cond:reach}
                \end{align}
            \end{subequations}
        \end{defn}

        Intuitively, condition \eqref{cond:deriv} requires that $H$ does not increase within the safe set excluding the target set. Combined with the boundary condition \eqref{cond:boundary}, this ensures that trajectories starting from the 0-sublevel set of $H$ will never leave $\Safe$ and will eventually reach the target set. Unlike traditional control barrier functions, which require \eqref{cond:deriv} to be satisfied only at the boundary of $\Safe$, we extend this requirement to a larger region. This is essential to ensure that trajectories can eventually enter the target set.

        Condition \eqref{cond:reach} is employed to exclude trajectories that would stay in $\overline{\Safe \setminus \Target}$ indefinitely. By integrating both sides of \eqref{cond:reach} over a sufficiently long time, the continuous differentiability of $w(\cdot)$ and the closeness of $\overline{\Safe \setminus \Target}$ will cause the right side of \eqref{cond:reach} to approach $0$, thereby excluding such trajectories from the 0-sublevel set of $H$. In fact, if we replace $w(\cdot)$ with $\frac{1}{\beta}H(\x_t)$ in condition \eqref{cond:reach}, with $\beta>0$, we derive a condition of the form $\beta H(\x_t)\ge \frac{d H(x_t)}{dt}$, which aligns with the conventional control barrier functional conditions \cite{ames2020integral}. Therefore, Definition \ref{df:rabfal} is more expressive and likely to produce a less conservative inner-approximation of the reach-avoid set.

        Now, we present the theorem for inner-approximating the reach-avoid set of DDEs. 
%

        \begin{thm}
        \label{thm:inner-approx-ra}
            Given a DDE of the form \eqref{eq:dde}, safe set $\Safe$, and target set $\Target$, if $H(\cdot):\ContiFun{[-\tau,0]}{\domain}\to\Real$ is a RABF, then the set $\RA_{in}$, defined by the 0-sublevel set of $H$, i.e.,
            \begin{align}\label{eq:ra-in}
                \RA_{in} \DefSym \{\pphi\in\ContiFun{[-\tau,0]}{\Safe} \mid H(\pphi) < 0\}
            \end{align}
            is an inner-approximation of $\RA$.
        \end{thm}
        \begin{pf}
            (i) First, we prove that trajectories originating from $\RA_{in}$ will remain within the safe set. From condition \eqref{cond:deriv} and the fact that $\RA_{in}$ is a 0-sublevel set of $H(\cdot)$, for any trajectory $\x^{\pphi}(\cdot)$ of \eqref{eq:dde} with $\pphi\in\RA_{in}$, the following inequality holds:
            \begin{align}\label{eq:reach}
                H(\x^{\pphi}_t) \le H(\pphi) < 0 \qquad \forall t\in\PosReal,
            \end{align}
            implying that the trajectories will never touch the boundary of $\Safe$ by condition \eqref{cond:boundary}.
    
            (ii) Next, we prove that trajectories originating from $\RA_{in}$ always reach the target set $\Target$ in finite time. Assume there exists $\pphi\in\RA_{in}$ such that $\x^{\pphi}(t)\in\overline{\Safe\setminus\Target}$ for all $t\in\PosReal$. Integrating both sides of \eqref{cond:reach} from $0$ to $p$ ($p>0$), we get $\int_0^p H(\x^{\pphi}t) dt \ge \int_{0}^{p} dw(\x^{\pphi}(t))$. From \eqref{eq:reach}, we conclude that $pH(\pphi)\ge \int_0^p H(\x^{\pphi}t) dt$, therefore:
            \begin{align*}
                H(\pphi) \ge \frac{w(\x^{\pphi}(p)) - w(\pphi(0))}{p}.
            \end{align*}
            Because $w(\cdot)$ is continuously differentiable and $\overline{\Safe\!\setminus\!\Target}$ is closed, $w(\cdot)$ is bounded. Therefore, as $p$ approaches infinity (since $\x^{\pphi}(\cdot)$ always remains in $\Safe\!\setminus\!\Target$), we derive that $H(\pphi)\ge 0$, which contradicts the condition that $\pphi\in\RA_{in}$.
            \qed
        \end{pf}

        The constraints in Definition~\ref{df:rabfal} are generally unsolvable. Therefore, in the next subsection, we will relax these constraints to make the synthesis of RABFs computable.

    \subsection{Inner-approximating Reach-Avoid Set}\label{subsec:inner-ra}
        To efficiently inner-approximate the reach-avoid set using RABFs, consider the RABFs defined in Definition~\ref{df:rabfal} with the following form:
        \begin{equation}\label{eq:rabfal}
            H(\x_t) = h_0(\x_t(0)) + \int_{-\tau}^{0}h_1(\x_t(\theta))d\theta
        \end{equation}
        where $h_0(\cdot):\Realn\to\Real$ and $h_1(\cdot):\Realn\to\Real$ are continuous differential.

        This form of RABFs is defined for two reasons.
        \begin{enumerate*}[label=(\roman*)]
            \item To incorporate the historical values of DDE trajectories, which is essential for RABFs in the context of DDEs, necessitating the integration term. 
            \item This form of RABFs is versatile, it can be easily reformulated to achieve a reach-avoid set in $\Realn$ by setting $h_1(\cdot)$ to 0.
        \end{enumerate*}
        Regarding RABFals with the form \eqref{eq:rabfal}, we have the following theorem. 
        \begin{thm}
            \label{lm:relx-inner}
            Given a DDE of form \eqref{eq:dde}, a safe set $\Safe$ and a target set $\Target$, if there exists a RABFal $H(\cdot)$ with the form \eqref{eq:rabfal}, and a continuous differentiable function $w:\domain\to\Real$, such that
            \begin{subequations}
                \begin{align}
                    &\frac{\partial h_0(\x)}{\partial \x}\cdot \f(\x,\x) + \tau \e_1(\x)^T\cdot C  \le 0, \forall \x\in\overline{\Safe\!\setminus\!\Target}, \label{cond:deriv-relx}\\
                    &h_0(\x) + \tau h_1(\x) + \frac{\tau^2}{2} \e_2(\x)^T\cdot C  \ge 0, 
                    \forall \x\in\partial\Safe,\label{cond:boundary-relx}\\
                    &\begin{aligned}
                        &h_0(\x) + \tau h_1(\x) + \frac{\tau^2}{2}\e_2(\x)^T\!\cdot\! C - \tau \e_3(\x)^T\!\cdot C\\
                        &~-\frac{\partial w(\x)}{\partial \x}\cdot\f(\x,\x) \ge 0, 
                        \forall \x\in \overline{\Safe\!\setminus\!\Target}, \label{cond:reach-relx}
                    \end{aligned}
                \end{align} 
            \end{subequations}
            where $\bm{e}_1,\bm{e}_2,\bm{e}_3\in\Poly{\x}^n$ and $\bm{C}\in\Realn$ are auxiliary polynomials satisfying (the inequalities are coordinate-wise)
                \begin{subequations}\label{eq:auxi}
                    \begin{align}
                        &\bm{e}_1(\x)^T  \ge \left|\frac{\partial h_1(\y)}{\partial\y}-\frac{\partial h_0(\x)}{\partial \x}\cdot\frac{\partial \f(\x,\y)}{\partial \y}\right|,\,\forall \x,\y\in\overline{\Safe\!\setminus\!\Target},\label{eq:e1} \\
                        &\bm{e}_2(\x)^T  \ge \left|\frac{\partial h_1(\x)}{\partial \x}\right|,\, \forall \x\in\overline{\Safe\!\setminus\!\Target},\label{eq:e2} \\
                        &\bm{e}_3(\x)^T  \ge \left|\frac{\partial w(\x)}{\partial \x}\cdot\frac{\partial \f(\x,\y)}{\partial \y}\right|,\, \forall \x,\y\in\overline{\Safe\!\setminus\!\Target},\label{eq:e3}\\
                        &\bm{C}  \ge \left|\f(\x,\y)\right|,\, \forall \x,\y\in\overline{\Safe\!\setminus\!\Target},\label{eq:c}
                    \end{align}
                \end{subequations}

             \noindent
             then $\RA_{in}$ defined by \eqref{eq:ra-in} is an inner-approximation of the reach-avoid set.
        \end{thm}
  
        Intuitively, Theorem~\ref{lm:relx-inner} aims to use the auxiliary polynomials in \eqref{eq:auxi} to provide a more precise bound for the delay terms of $\x_t(\cdot)$ (i.e., $\x_t(\theta)$ for $\theta \in [-\tau,0)$). This approach offers a more accurate bound compared to the $n$-dimensional ball method used in \cite{xue2021reach} and the method in \cite{prajna2005methods}, which treats the historical terms as individual variables.

        The correctness of Theorem~\ref {lm:relx-inner} follows directly from applying Theorem~\ref {thm:inner-approx-ra} to RABF with the form \eqref{eq:rabfal} (cf. Cond. \eqref{cond:deriv}--\eqref{cond:boundary} and Cond. \eqref{cond:deriv-relx}--\eqref{cond:reach-relx}).  Cond. \eqref{eq:e1}--\eqref{eq:e3} are auxiliary conditions to control the extra terms during simplification. 

        \begin{pf} 
            We will show that Cond. \eqref{cond:deriv-relx}--\eqref{cond:reach-relx} imply Cond. \eqref{cond:deriv}--\eqref{cond:boundary} under auxiliary conditions \eqref{eq:e1}--\eqref{eq:e3}. Therefore, by Thm.\nobreakspace \ref {thm:inner-approx-ra}, $\RA_{in}$ is an inner-approximation of the reach-avoid set.

            First, we present two equalities from \cite{fridman2008input}, which rewrite the delay term $\x(t-\tau)$ as an integral,
            \begin{align}
                &\begin{aligned}
                    &\f(\x(t),\x(t-\tau)) = \f(\x(t),\x(t))\\
                    &\qquad\qquad\qquad\quad -\int_{t-\tau}^{t}\frac{\partial \f(\x(t),\x(\theta))}{\partial \x(\theta)}\cdot \dot\x(\theta) d\theta,
                \end{aligned}\label{eq:fx_tau}\\
                &h\left(\x(t-\theta )\right) = h\left(\x(t)\right)-\int_{t-\theta}^{t}\frac{\partial h}{\partial \x} \cdot \dot \x(\rho) d\rho, \label{eq:x_tau}
            \end{align}
            where $\theta >0$, and $h(\cdot)$ is any continuous differentiable function.

            We now show that Cond. \eqref{cond:deriv} holds under Cond. \eqref{cond:deriv-relx}. Since $H(\cdot)$ takes the form of \eqref{eq:rabfal}, we have
            \begin{align*}
                \frac{dH(\x_t)}{dt} 
                = & \frac{\partial h_0(\x(t))}{\partial \x(t)}\f(\x(t),\x(t)) + \int_{t-\tau}^{t}\left( \frac{\partial h_1(\x(\theta))}{\partial \x(\theta)}-\right.\\
                &\left.\frac{\partial h_0(\x(t))}{\partial \x(t)}\cdot\frac{\partial \f(\x(t),\x(\theta))}{\partial \x(\theta)} \right)\cdot \dot \x(\theta)d\theta\TAG{by Eq.\nobreakspace \textup {(\ref {eq:fx_tau})}}\\
                \le& \frac{\partial h_0(\x(t))}{\partial \x(t)}\f(\x(t),\x(t)) + \int_{t-\tau}^{t}\left| \frac{\partial h_1(\x(\theta))}{\partial \x(\theta)}-\right.\\
                &\left.\frac{\partial h_0(\x(t))}{\partial \x(t)}\cdot\frac{\partial \f(\x(t),\x(\theta))}{\partial \x(\theta)} \right|\cdot \left| \dot \x(\theta)\right| d\theta\\
                \le& 
                \frac{\partial h_0}{\partial \x(t)} \f(\x(t),\x(t)) + \bm{e}_1(\x(t))^T\cdot\tau \bm{C} \TAG{by auxiliary conditions \eqref{eq:e1}, \eqref{eq:c}}\\
                \le& 0. \TAG{by Cond. \eqref{cond:deriv-relx}}
            \end{align*}
            This proves Cond. \eqref{cond:deriv}.

            Next, we prove that Cond. \eqref{cond:boundary} holds under Cond. \eqref{cond:boundary-relx}. Since $H(\cdot)$ takes the form  \eqref{eq:rabfal}, we have
            \begin{align*}
                &H(\x_t)\\ 
                =& h_0(\x(t)) + \int_{-\tau}^0 \left(h_1(\x(t))-\int_{t+\theta}^{t}\frac{\partial h_1(\x(\rho))}{\partial \x(\rho)}\cdot\dot\x(\rho)d\rho \right)d\theta\TAG{by Eq.\nobreakspace \textup {(\ref {eq:fx_tau})}}\\
                =& h_0(\x(t)) + \tau h_1(\x(t))-\int_{-\tau}^0 \int_{t+\theta}^{t}\frac{\partial h_1(\x(\rho))}{\partial \x(\rho)}\cdot\dot\x(\rho)d\rho d\theta\\
                \ge& h_0(\x(t)) + \tau h_1(\x(t)) - \int_{-\tau}^0 \int_{t+\theta}^{t}\left|\frac{\partial h_1(\x(\rho))}{\partial \x(\rho)}\right|\cdot\left|\dot\x(\rho)\right|d\rho d\theta\\
                \ge& h_0(\x(t)) + \tau h_1(\x(t)) - \int_{-\tau}^{0}\int_{t+\theta}^{t}\bm{e}_2(\x(t))^T\cdot \bm{C} \,d\rho d\theta\TAG{by auxiliary conditions \eqref{eq:e2}, \eqref{eq:c}}\\
                =& h_0(\x(t)) + \tau h_1(\x(t)) + \frac{\tau^2}{2}\bm{e}_2(\x(t))^T\cdot \bm{C}\\
                \ge& 0. \TAG{by Cond. \eqref{cond:boundary-relx}}  
            \end{align*}
            Thus, Cond. \eqref{cond:boundary} holds.
        
            Finally, we show that Cond. \eqref{cond:reach} holds under Cond. \eqref{cond:reach-relx}. In fact, 
            \begin{align*}
                &H(\x_t)-\frac{dw(\x_t(0))}{dt}\\
                =& h_0(\x(t)) + \tau h_1(\x(t))-\int_{-\tau}^0 \int_{t+\theta}^{t}\frac{\partial h_1(\x(\rho))}{\partial \x(\rho)}\cdot\dot\x(\rho)d\rho d\theta\\
                &+\int_{t-\tau}^{t}\frac{\partial w(\x(t))}{\partial \x(t)}\cdot\frac{\partial\f(\x(t),\x(\theta))}{\partial\x(\theta)}\cdot \dot \x(\theta)\,d\theta\\
                &-\frac{\partial w(\x(t))}{\partial \x(t)}\cdot\f(\x(t),\x(t)) \TAG{by Eq.\nobreakspace \textup {(\ref {eq:fx_tau})} and\nobreakspace  \textup {(\ref {eq:x_tau})}}\\
                \ge& h_0(\x(t)) + \tau h_1(\x(t)) + \frac{\tau^2}{2}\e_2(\x(t))^T\cdot C\\
                & -\frac{\partial w}{\partial \x(t)}\f(\x(t),\x(t))-\int_{t-\tau}^{t}\e_3(\x(t))\cdot C\,d\theta\TAG{by auxiliary conditions \eqref{eq:e2}, \eqref{eq:e3}, \eqref{eq:c}}\\
                =& h_0(\x(t)) + \tau h_1(\x(t)) + \frac{\tau^2}{2}\e_2(\x(t))^T\!\!\cdot\! C
                -\tau\e_3(\x(t))\!\cdot\! C\\
                &-\frac{\partial w}{\partial \x(t)}\f(\x(t),\x(t))
                \ge 0. \TAG{by Cond. \eqref{cond:reach-relx}}
            \end{align*}
            Thus, Cond. \eqref{cond:reach} holds. This completes the proof. \qed
        \end{pf}

        When $\f$ in DDE \eqref{eq:dde} is polynomial in $\x(t)$ and $\x(t-\tau)$, and both the safe set and target set are defined by polynomials (i.e., $\fsafe\in\Poly{\x}$, $\ftarget\in\Poly{\x}$), all constraints in Theorem~\ref {lm:relx-inner} can be reduced to an SDP problem. Therefore, we can use SDP solving techniques \cite{prajna2005sostools} to synthesize RABF $H(\x_t)$ of the form \eqref{eq:rabfal} in a standard way. Next, we will demonstrate the advantage of our method compared to the leading approach proposed in \cite{xue2021reach} through an example.

        \begin{exmp}\label{exp:inRA}
            Consider the DDE defined by the vector field in mode $q_1$ of Example\nobreakspace\ref{exp:running}: 
            \begin{align*}
                &\qquad\dot x_1(t) = 0.5 x_2(t) + 0.5 x_2(t-\tau),\\
                &\qquad\dot x_2(t) = -0.5 x_1(t) - 0.5 x_1(t-\tau) - 1.5 x_2(t).
            \end{align*}
            We will compare our method with the approach presented in \cite{xue2021reach} in terms of conservatism, computation time, and the size of $\tau$.
            \begin{figure}[h]
                \begin{minipage}{0.25\textwidth}
                    \hspace{0.2cm}
                    \begin{adjustbox}{max width=1\linewidth}
                        \begin{tikzpicture}
                            \draw (0, 1.8) node[inner sep=0] {\includegraphics[width=0.35\linewidth]{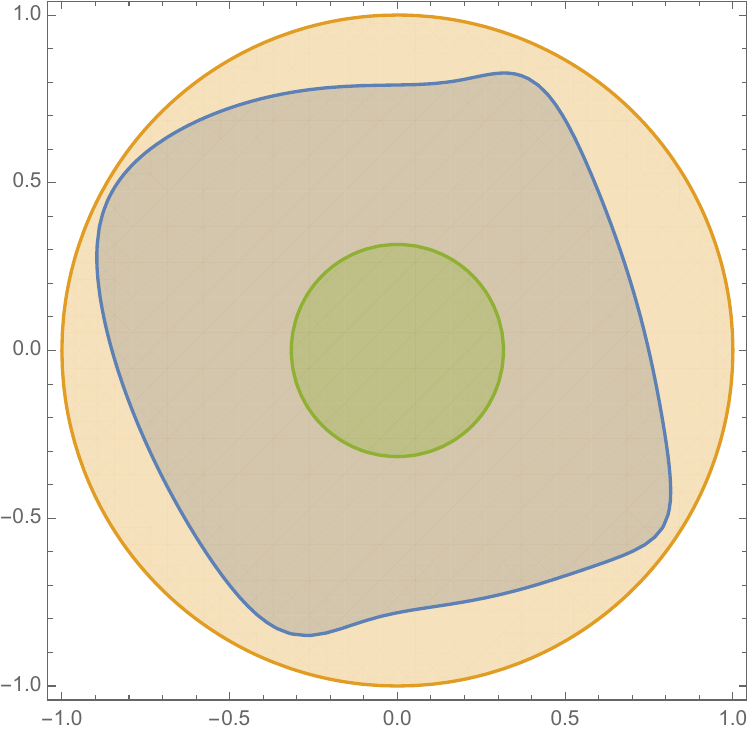}};
                            \node at(0,0.93) {\tiny{Our Method}};
                            \draw (0, 0) node[inner sep=0] {\includegraphics[width=0.35\linewidth]{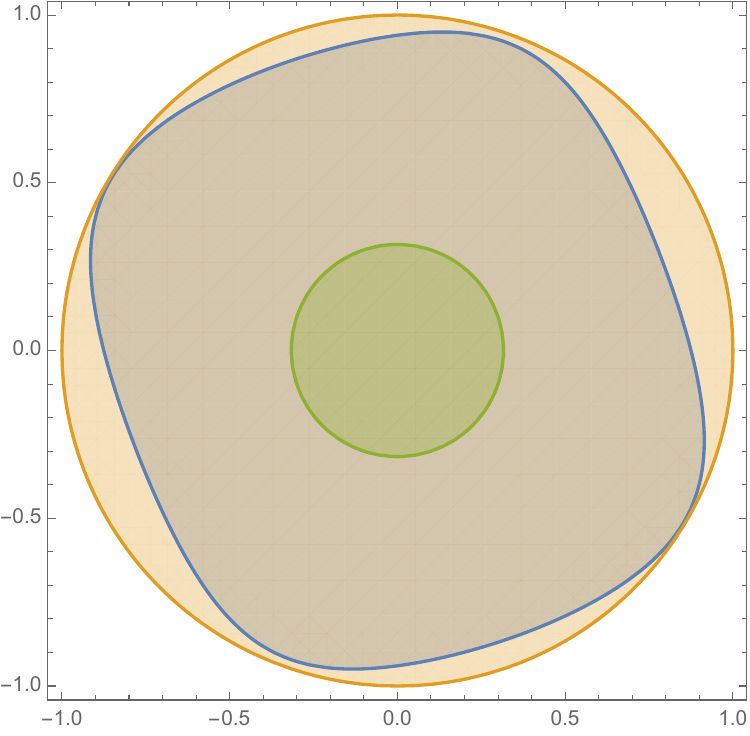}};
                            \node at(0,-0.9) {\tiny{\cite{xue2021reach}'s Method}};
                        \end{tikzpicture}
                    \end{adjustbox}
                \end{minipage}
                \hspace{-2cm}
                \begin{minipage}{0.25\textwidth}
                    \begin{adjustbox}{max width=1.2\linewidth}
                        \begin{tabular}{c c cc c cc}
                        \toprule
                        \multirow{2}{*}{$\tau$}&~ &\multicolumn{2}{c}{\textsf{Our Method}} &~& \multicolumn{2}{c}{\textsf{\cite{xue2021reach}'s Method}}\\
                        \cmidrule{3-4} \cmidrule{6-7}
                        & & Validity & Time & ~ & Validity & Time \\
                        \midrule
                        0.1 && \cmark &  5.81s &&  \cmark & 4.09s \\
                        0.2 && \cmark &  5.24s &&  \cmark & 4.73s \\
                        0.3 && \cmark &  5.28s &&  \xmark & 4.25s \\
                        0.4 && \cmark &  6.10s &&  \xmark & 3.64s \\
                        0.5 && \cmark &  5.91s &&  \xmark & 3.77s \\
                        0.6 && \cmark &  5.05s &&  \xmark & 3.78s \\
                        \bottomrule
                        \end{tabular}
                    \end{adjustbox} 
                \end{minipage}
                \caption{Comparison between the method in \cite{xue2021reach} and ours}
                \label{fig:in-ra}
            \end{figure}
            
            The figures on the left side of Fig. \ref{fig:in-ra} compare the conservatism of two methods when $\tau=0.1$. The blue regions represent the inner-approximated reach-avoid set\footnote{We constrained the functions in $\RA_{in}$ calculated by our method to constant functions for visualization.}. The orange region denotes the safe set, and the green region represents the target set. Our method exhibits greater conservatism compared to the existing approach, primarily due to the additional constraints introduced by the auxiliary polynomials.

            The table on the right side of Fig. \ref{fig:in-ra} displays computation times and result validity for various $\tau$ values. Validity is determined by the emptiness of the inner-approximated reach-avoid sets\footnote{This can be verified using Satisfiability Modulo Theories (SMT) tools such as Z3 \cite{de2008z3}}. Our method demonstrates superior capability in handling DDEs with larger delays $\tau$ compared to \cite{xue2021reach}, owing to its more precise bound for the delay terms. However, our method requires more computation time due to the incorporation of more complex constraints.
        \end{exmp}
        
        In summary, the reach-avoid inner-approximation method we propose in this section serves as an alternative to the existing approach and demonstrates greater efficiency in handling reach-avoid problems for DDEs with larger delays.

\section{Constructing Discrete Directed Graph}\label{sec:consDG}
  In this section, we introduce the second step of our approach, that is, to construct a discrete directed graph from the considered  $\edHAabb$ $\mathcal{H}$ based on the computed 
  reach-avoid sets for all its modes. 
  
   \subsection{Mode Partition}\label{subsec:split}
        To avoid the difficulty caused by 
        non-determinism in the third step of reset controller synthesis, 
        the constructed discrete directed graph from $\mathcal{H}$ should be 
        deterministic. So, 
        in this subsection, by employing Alg.\nobreakspace \ref {alg:inner-ra}, we will give a mode partition algorithm to split a single mode of $\edHAabb$ into some sub-modes, 
        such that for any two of these sub-modes the reach-avoid sets of their guard conditions and target sets are disjoint. 
        Moreover, such partition will help us to abstract away the continuous part of $\edHAabb$ and transform it into a discrete directed graph, which in turn simplifies the process of synthesizing reset maps (as will be discussed in Section \ref{sec:discrete}).  
        
        We firstly employ the running example to give an intuitive explanation of our partition algorithm. 
        
        \begin{exmp}\label{exp:mode_seg}
            Consider Example\nobreakspace \ref {exp:running}, where the safe set defined as $\Safe_{q_1}=\Safe_{q_2}=\Safe_{q_3}=\{(x_1,x_2)\mid x_1^2+x_2^2 -1\le 0\}$, the target set is $\Target_{q_3}=\{(x_1,x_2)\mid x_1^2+x_2^2\le 0.1\}$ and $\Target_{q_1}=\Target_{q_2}=\emptyset$. The procedure of mode partition for $q_1$ is illustrated in Fig.\nobreakspace \ref {fig:seg_q1}.
            We denote the reach-avoid set computed by Alg.\nobreakspace \ref {alg:inner-ra} in mode $q_i$ as $\RA_{in}(i,j)$, where $g_{ij}$ represents the target (as defined in Fig.\nobreakspace \ref {fig:seg_q1}). Therefore, the regions enclosed by the green, red, and purple curves in mode $q_1$ (as shown in Fig.\nobreakspace \ref {fig:seg_q1}) can be described as follows:
            \begin{align*}
                &\textcolor{teal}{\RA_{in}(1,1)\DefSym \InnerRA{\f_1}{\Dom(q_1)\cap\Safe_{q_1}}{g_{11}\setminus\Dom(q_1)}},\\
                &\textcolor{purple}{\RA_{in}(1,2)\DefSym\InnerRA{\f_1}{\Dom(q_1)\cap\Safe_{q_1}}{g_{12}\setminus\Dom(q_1)}},\\
                &\textcolor{violet}{\RA_{in}(1,3)\DefSym \InnerRA{\f_1}{\Dom(q_1)\cap\Safe_{q_1}}{g_{13}\setminus\Dom(q_1)}}.
            \end{align*} 
            Please note that, strictly speaking, $\RA_{in}(\cdot)$ represents a set of functions and should not be depicted in the same figure with $\Dom(q_1)\in\Real^2$. However, for the purpose of providing an intuitive perspective, we included these sets together.
            \begin{figure}[htb]
                \centering
                \vspace*{-4mm}
                \begin{adjustbox}{max width = 1\linewidth}
                    \scalebox{1.4}{
                        \begin{tikzpicture}[scale=1,
                            pattern1/.style={draw=black,pattern color=black!60,pattern=north east lines},
                            pattern2/.style={draw=black,pattern color=cyan!60,pattern=north east lines},
                            pattern3/.style={draw=black,pattern color=orange,pattern=north east lines}
                            ]
                            \draw (0,0) node {\includegraphics[width=10cm]{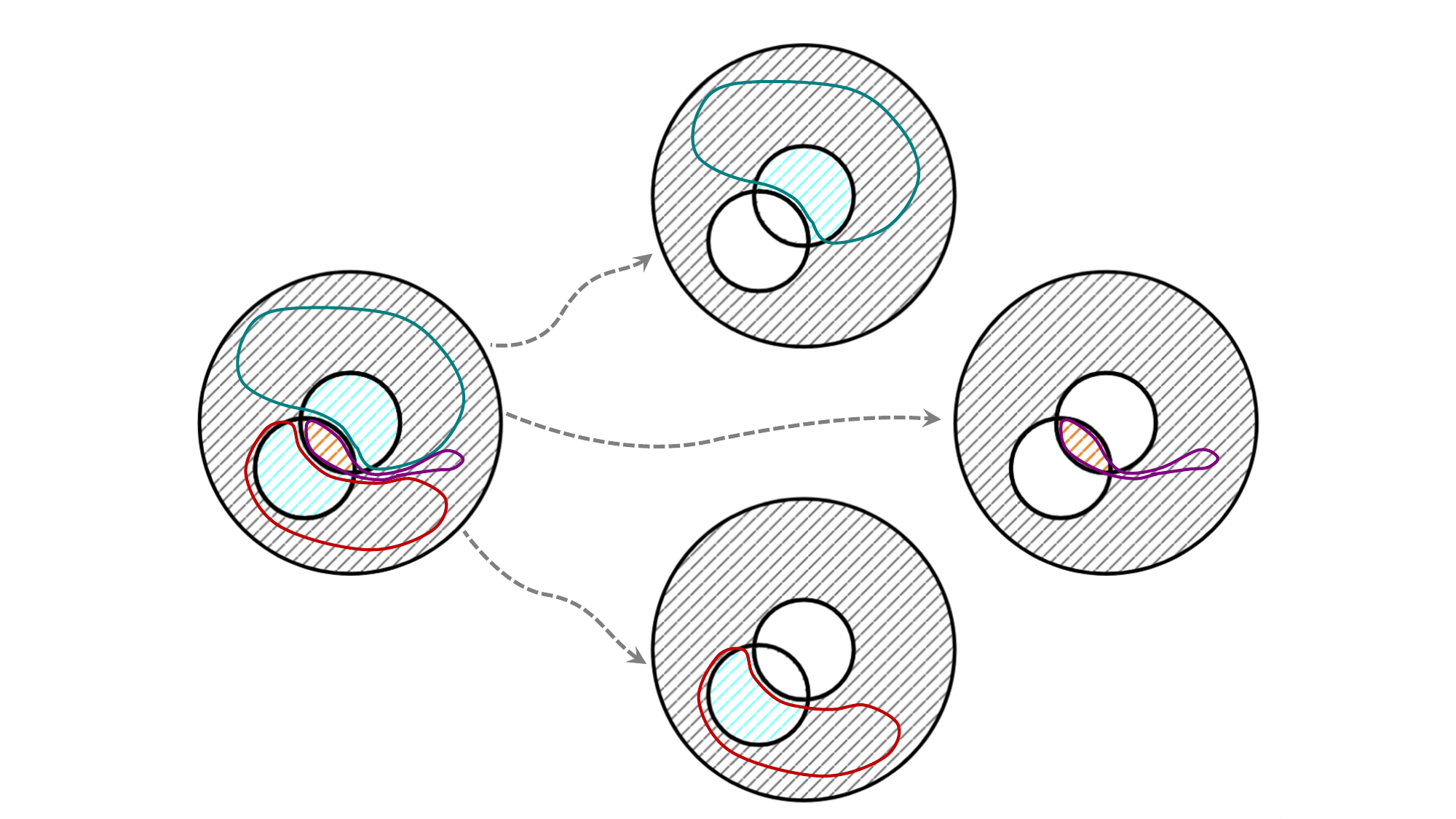}};

                            \node at(-2.6,1.1) {\tiny{$q_1$}};
                            \node at(-2.4,0.5) {\tiny{$\Dom(q_1)$}};
                            \node at(-2.6,0.04) {\tiny{$\Guard(\!e_1\!)$}};
                            \node at(-2.93,-0.55) {\tiny{$\Guard(\!e_2\!)$}};

                            \node at(0.5,2.65) {\tiny{$q_{11}$}};
                            \node at(0.8,2) {\tiny{$\Dom(q_1)$}};
                            \node at(-1.6,1.56) {\tiny{$g_{11}:=\Guard(\!e_1\!)\!\!\setminus\!\!\Guard(\!e_2\!)$}};
                            \draw[->,cyan] (-1.2,1.7) to[bend left=35] (0.4,1.6);

                            \node at(0.5,-0.5) {\tiny{$q_{12}$}};
                            \node at(0.8,-1.1) {\tiny{$\Dom(q_1)$}};
                            \node at(-1.6,-2) {\tiny{$g_{12}:=\Guard(\!e_2\!)\!\!\setminus\!\!\Guard(\!e_1\!)$}};
                            \draw[->,cyan] (-1.2,-2.15) to[bend right=35] (0,-2);

                            \node at(2.5,1.1) {\tiny{$q_{13}$}};
                            \node at(2.8,0.5) {\tiny{$\Dom(q_1)$}};
                            \node at(2.7,-1.4) {\tiny{$g_{13}:=\Guard(\!e_2\!)\!\cap\!\Guard(\!e_1\!)$}};
                            \draw[->,orange] (2.5,-1.25) to[bend left=35] (2.5,-0.3);

                        \end{tikzpicture}
                   }
                \end{adjustbox}
                \caption{Mode partition of $q_1$. On the left side, we have mode $q_1$ with guard conditions $\Guard(e_1)$ and $\Guard(e_2)$ represented by blue slashes, and their intersection depicted by orange slashes. The reach-avoid set to $\Guard(e_1) \cup \Guard(e_2)$ can be partitioned into three disjoint regions: $g_{11}$, $g_{12}$, and $g_{13}$, as shown above. Accordingly, mode $q_1$ is partitioned into three sub-modes: $q_{11}$, $q_{12}$, and $q_{13}$.}
                \label{fig:seg_q1}
            \end{figure}
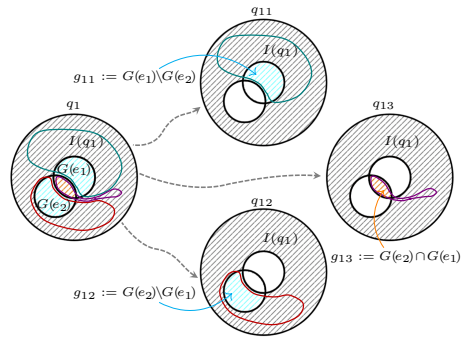
             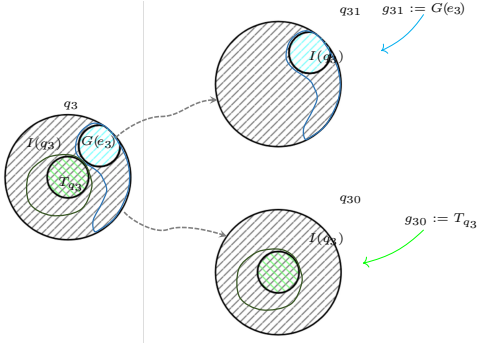
\begin{figure}[h]
                \vspace*{-1mm}
                 \centering
                 \begin{adjustbox}{max width = 1\linewidth}
                     \scalebox{1.4}{
                         \begin{tikzpicture}[scale=1,
                            pattern1/.style={draw=black,pattern color=black!60,pattern=north east lines},
                            pattern2/.style={draw=black,pattern color=cyan!60,pattern=north east lines},
                            pattern3/.style={draw=black,pattern color=green!60,pattern=north west lines}
                            ]
                            \draw (0,0) node {\includegraphics[width=10cm]{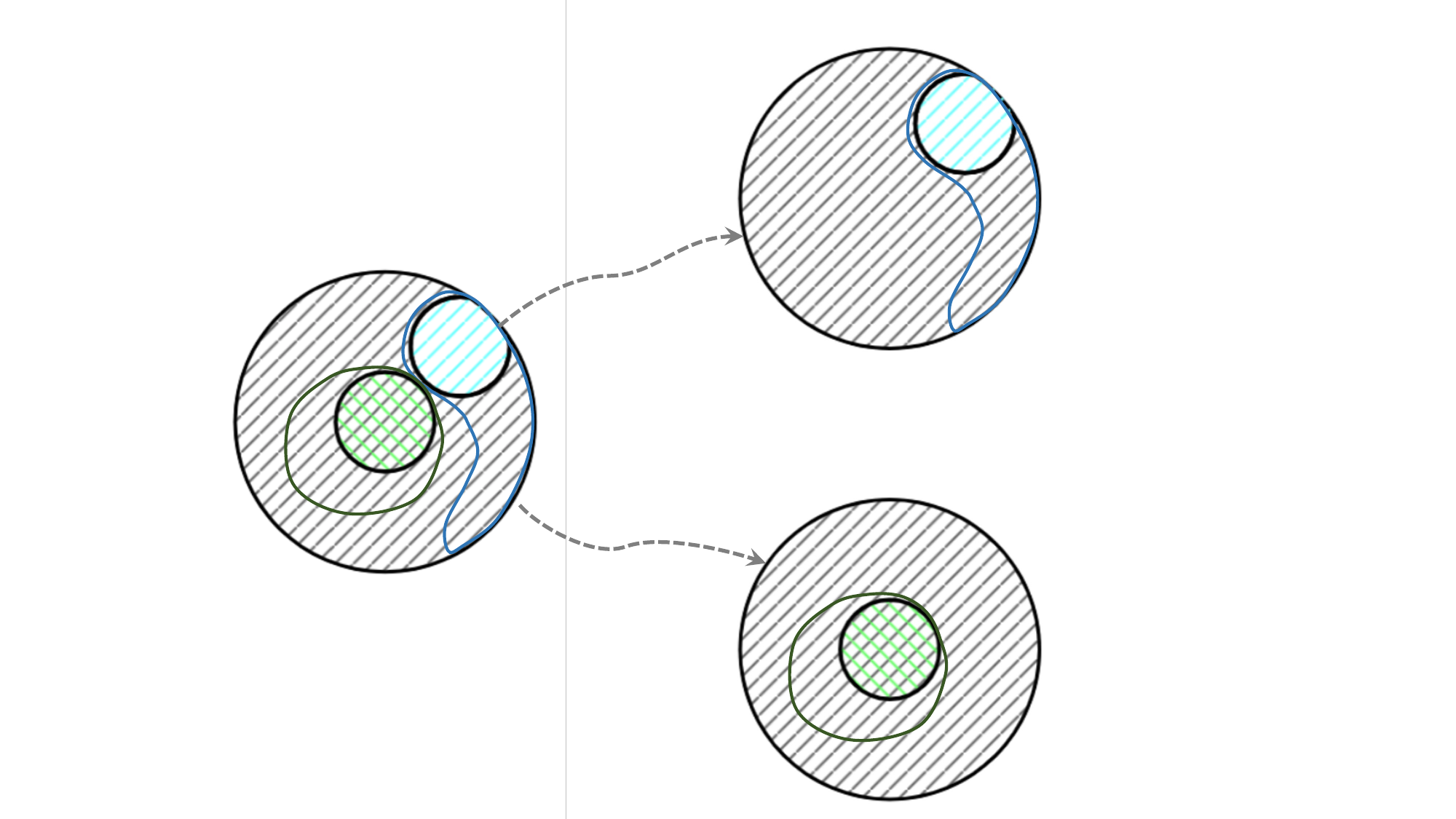}};

                            \node at(-2.3,1.1) {\tiny{$q_3$}};
                            \node at(-2.75,0.47) {\tiny{$\Dom(q_3)$}};
                            \node at(-1.85,0.5) {\tiny{$\Guard(\!e_3\!)$}};
                            \node at(-2.3,-0.2) {\tiny{$\Target_{q_3}$}};

                            \node at(2.3,2.65) {\tiny{$q_{31}$}};
                            \node at(1.9,1.9) {\tiny{$\Dom(q_3)$}};
                            \node at(3.5,2.7) {\tiny{$g_{31}:=\Guard(\!e_3\!)$}};
                            \draw[->,cyan] (3.5,2.6) to[bend left = 15] (2.8,2);

                            \node at(2.3,-0.45) {\tiny{$q_{30}$}};
                            \node at(1.9,-1.1) {\tiny{$\Dom(q_3)$}};
                            \node at(3.8,-0.8) {\tiny{$g_{30}:=\Target_{q_3}$}};
                            \draw[->,green] (3.5,-0.95) to[bend left = 15] (2.5,-1.5);
                            %
                            %
                         \end{tikzpicture}
                    }
                 \end{adjustbox}
                \caption{Mode partition of $q_3$. The left side is mode $q_3$ with the guard condition $\Guard(e_3)$ (blue slashes) and the target set $\Target_{q_3}$ (green slashes). 
                Correspondingly, $q_1$ is  partitioned into two sub-modes: $q_{30}$ with $g_{30}=\Target_{q_3}$, and $q_{31}$ with $g_{31}=\Guard(e_3)$.}
                \label{fig:seg_q3}
             \end{figure}
            
            The partition of $q_3$ is illustrated in Fig.\nobreakspace \ref {fig:seg_q3}. We denote the regions enclosed by the green and blue curves, computed using Alg.\nobreakspace \ref {alg:inner-ra} with $g_{30}$ and $g_{31}$ as the respective targets, as $\RA_{in}(3,0)$ and $\RA_{in}(3,1)$.

            After partitioning all modes, new edges should be introduced. Let's consider  $q_{31}$, edges  from  $q_{31}$  to
            the sub-modes of $q_1$ are introduced, i.e., including  $(q_{31},q_{11}),(q_{31},q_{12})$, and $(q_{31},q_{13})$. 
            Accordingly, the reset maps associated with them 
            are given as: 
            \begin{itemize}
                \item $\Reset^m\left((q_{31},q_{11}),\Guard^m(q_{31},q_{11})\right)=\RA_{in}(1,1)$,
                \item $\Reset^m\left((q_{31},q_{12}),\Guard^m(q_{31},q_{12})\right)=\RA_{in}(1,2)$,
                \item $\Reset^m\left((q_{31},q_{13}),\Guard^m(q_{31},q_{13})\right)= \RA_{in}(1,3)$.
            \end{itemize} 
            where  $\Guard^m(q_{31},q_{11})\!=\!\Guard^m(q_{31},q_{12})\!=\!\Guard^m(q_{31},q_{13})\!=\!g_{31}$.
            
            Similarly, for $q_{13}$, edges from  $q_{13}$ 
            to the sub-modes of $q_2$ and $q_3$ should  be introduced, 
            whose reset map can be defined in the same way.
        \end{exmp}

        \begin{rem}
           Note that  we set the third input of $\InnerRA{}{}{}$ in $\RA_{in}(1,1)$ to $g_{11}\!\setminus\!\Dom(q_1)$ rather than $g_{11}$ simply in order to remove 
             non-determinism 
             between continuous evolution and 
             discrete transition. As an execution remains in $g_{11}\cap\Dom(q_1)$, it can either  
             take the discrete transition immediately, or continue 
             the continuous evolution. However, whenever it reaches to 
             $g_{11}\cap \neg \Dom(q_1)$, the discrete jump has to happen. 
        \end{rem}
        
        We present the mode partition algorithm in Alg.\nobreakspace \ref {alg:mode-seg}, where $\InnerRA{\f_i}{\Safe_{q_i}}{\Target_{q_i}}$ denote the reach-avoid analysis procedure presented in Section \ref{sec:continuous}, with the dynamic being $\f_i$, the safe set being $\Safe_{q_i}$, the target set being $\Target_{q_i}$. The algorithm proceeds as follows: 
        Firstly, it iteratively partitions all modes of $\edHAsyb$ (line \ref{alg2:mode-start} to line \ref{alg2:mode-end}). During each iteration, a sub-mode is added if the reach-avoid set to the target set is non-empty (line \ref{alg2:target-start} to line \ref{alg2:target-end}). 
        Next, the guard conditions are partitioned so that the corresponding reach-avoid sets  are disjoint (line \ref{alg2:guard-start}). 
        Based on the resulting partition, each mode is also partitioned 
        into sub-modes accordingly (line \ref{alg2:g_s} to line \ref{alg2:guard-end}).
        Finally, we introduce edges to $\edHAsyb^m$ in accordance with $\edHAsyb$ and synthesize a reset map accordingly (line \ref{alg2:edge-start} to line \ref{alg2:edge-end}).
        
        The correctness of Alg.\nobreakspace \ref {alg:mode-seg} is guaranteed  by Thm.\nobreakspace \ref {lm:relx-inner}.
    \begin{small}
        \begin{algorithm*}[htb]
            \caption{$\ModeSeg{\edHAsyb}{\Safe}{\Target}$}
            \label{alg:mode-seg}
            \begin{multicols}{2}
            \begin{algorithmic}[1]
                \REQUIRE  A $\edHAabb$ $\edHA$, a safe set $\Safe\subseteq \DisState\times\ContiState$, a target set $\Target\subseteq\DisState\times\ContiState$
                \ENSURE  The $\edHAabb$ $\edHAsyb^m=(\DisState^m,\ContiState,\Dom^m,\ContiDyn^m,\Init^m,\DisJump^m,\allowbreak\Guard^m,\STime^m,\Reset^m)$ and target set $\Target^m$, safe set $\Safe^m$ after partition\\
                /*~\textit{Partition modes of $\edHAsyb$ into sub-modes based on the result of reach-avoid analysis}~*/
                \FOR {$q_i\in\DisState$}\label{alg2:mode-start}
                    \STATE $\RA_{in}(i,0)\gets\InnerRA{\f_i}{\Dom(q_i)\cap \Safe_{q_i}}{\Target_{q_i}}$\label{alg2:target-start}
                    \IF{$\RA_{in}(i,0)$ is not empty}
                        \STATE Add $q_{i0}$ to $\DisState^m$, with $\Dom^m(q_{i0})=\Dom(q_i)$, $\f_{i0}^m=\f_i$, $\Init^m(q_{i0})=\Init(q_i)\cap\RA_{in}(i,0)$, $\Target^m_{q_{i0}}=\Target_{q_i}$, $\Safe^m_{q_{i0}}=\Safe_{q_i}$
                    \ENDIF\label{alg2:target-end}
                    \STATE Partition the guard conditions of edges jumping from $q_i$ into disjoint sets, and store these sets in $\Guard_i$\label{alg2:guard-start}
                    \STATE $j\gets 1$
                    \FOR{$g_{ij}\in\Guard_i$}\label{alg2:g_s}
                        \STATE $\RA_{in}(i,j) \gets\InnerRA{\f_i}{\Dom(q_i)\!\cap\!\Safe_{q_i}}{g_{ij}\!\setminus\!\Dom(q_i)}$
                        \IF{$\RA_{in}(i,j)$ is not empty}
                            \STATE Add $q_{ij}$ to $\DisState^m$, with $\Dom^m(q_{ij})=\Dom(q_i)$, $\f_{ij}^m=\f_i$, $\Init^m(q_{ij})=\Init(q_i)\cap\RA_{in}(i,j)$, $T^m_{q_{ij}}=\emptyset$, $\Safe^m_{q_{ij}}=\Safe_{q_i}$
                        \ENDIF
                        \STATE $j\gets j+1$
                    \ENDFOR\label{alg2:guard-end}
                \ENDFOR\label{alg2:mode-end}\\
                /*~\textit{Rebuild the edges of $\edHAsyb^m$}~*/
                \FOR {$q_i\in\DisState$}\label{alg2:edge-start}
                \FOR{each sub-mode $q_{ij}$ of $q_i$ with guard condition $g_{ij}$ in it}
                    \STATE $Post(q_{ij})\gets \{p\in\DisState\mid g_{ij}\subseteq \Guard(q_i,p)\}$\\
                    /*~\textit{$Post(q_{ij})$ contains the modes in $\edHAsyb$ such that $g_{ij}$ is the intersection of guard conditions of edges from $q_i$ to these modes}~*/
                    \FOR{$q_j\in Post(q_{ij})$}
                        \FOR{each sub-mode $q_{jk}$ of $q_j$}
                            \STATE Add edge $e=(q_{ij},q_{jk})$ to $\DisJump^m$, with\\ $\Guard^m(e)=g_{ij}$,$\Reset^m(e,\Guard^m(e))=\RA_{in}(j,k)$
                        \ENDFOR
                    \ENDFOR
                \ENDFOR
                \ENDFOR\label{alg2:edge-end}
                \STATE $\edHAsyb^m=(\DisState^m,\ContiState,\Dom^m,\ContiDyn^m,\Init^m,\DisJump^m,\Guard^m,\STime^m,\Reset^m)$
                \RETURN $\edHAsyb^m$, $\Target^m$, $\Safe^m$
            \end{algorithmic}
        \end{multicols}
        \end{algorithm*}
    \end{small}

  \subsection{Transforming to Discrete Directed Graph}\label{subsec:trans}
    For a given 
    $\edHAabb$ $\edHAsyb=(\DisState,\ContiState,\Dom,\ContiDyn,\Init,\DisJump,\Guard,\STime,\Reset)$, 
    let $\edHAsyb^m=(\DisState^m,\ContiState,\Dom^m,\ContiDyn^m,\Init^m,\DisJump^m,\Guard^m,\STime^m,\Reset^m)$ be the resulting $\edHAabb$ after applying  Alg.\nobreakspace \ref {alg:mode-seg} 
    to $\edHAsyb$. 
    Thus, a discrete directed graph $\DG=(V,E,V_0,V_T)$ can be defined as follows:
    \begin{itemize}
        \item $V=Q^m$ is the set of vertices;
        \item $E=\DisJump^m$ is the set of directed edges (arcs);
        \item $V_0=\{q\in Q^m\mid \Init^m(q)\neq\emptyset\}$ is the set of initial vertices;
        \item $V_T=\{q\in Q^m \mid \Target^m_q\neq\emptyset\}$ is the set of target vertices.
    \end{itemize}  
    
       Note that our definition above  slightly  
        differs from the standard definition of directed graph
        by allowing $V_0$ and $V_T$. 

    The following running example illustrates the above transformation.
    \begin{exmp}\label{exp:trans}
        Let's continue  the $\edHAabb$ given in Example\nobreakspace \ref {exp:running} (see 
        Fig.\nobreakspace \ref {fig:hldo}). In Example\nobreakspace \ref {exp:mode_seg}, $q_2$ is partitioned into three sub-modes, $q_3$ is partitioned into two sub-modes, while 
        $q_2$ does not need to be partitioned. The resulting discrete directed graph is depicted in Fig.\nobreakspace \ref {fig:nfa}, where  $V_0=\{q_{11},q_{12},q_{13}\}$  and  $V_T=\{q_{30}\}$.
    \qedT
    \end{exmp}
   \begin{figure}[h]
        \centering
       \oomit{  \begin{minipage}[b]{0.45\textwidth}
            \centering
            \begin{adjustbox}{max width=1\linewidth}
            \scalebox{1}{
                \begin{tikzpicture}[mode/.style={circle,draw=black,thick,
                    inner sep=0pt,minimum size=15mm},
                    target/.style={circle,draw=black,double,thick,
                    inner sep=0pt,minimum size=5mm},]
                    \node (q1) at (0,0) [mode] {};
                    \node (q2) at (3,0) [mode] {};
                    \node (q3) at (1.5,2.6) [mode] {};

                    \draw[->,thick] (q1) to (q2);
                    \node [below of=q3, xshift=0cm,yshift=-1.4cm] {\tiny{$e_1,\Guard(e_1)$}};
                    \node [below of=q3, xshift=0cm,yshift=-1.8cm] {\tiny$\Reset(e_1\!,\!(x_1\!,\!x_2)\!)$};
                    \draw[->,thick] (q3) to[bend right=35] (q1);
                    \node [below of=q3, xshift=-1.9cm,yshift=0.3cm] {\tiny{$e_3,\Guard(e_3)$}};
                    \node [below of=q3, xshift=-2.3cm,yshift=0cm] {\tiny{$\Reset(e_3\!,\!(x_1\!,\!x_2)\!)$}};
                    \draw[->,thick] (q1) to[bend right=35] (q3);
                    \node [below of=q3, xshift=0.7cm,yshift=-0.1cm] {\tiny{$e_2,\Guard(e_2)$}};
                    \node [below of=q3, xshift=0.8cm,yshift=-0.35cm] {\tiny{$\Reset(e_2\!,\!(x_1\!,\!x_2)\!)$}};

                    \node [below of=q1,yshift=1.05cm] {\tiny$\begin{array}{c}
                        q_1\\
                        \dot{\x}=\f_1\\
                        \Dom(q_1)\\
                        \Init(q_1)
                    \end{array}$};
                    \node [below of=q2,yshift=1.0cm] {\tiny$\begin{array}{c}
                        q_2\\
                        \dot{\x}=\f_2\\
                        \Dom(q_2)
                    \end{array}$};
                    \node [below of=q3,yshift=1.05cm] {\tiny$\begin{array}{c}
                        q_3\\
                        \dot{\x}= \f_3\\
                        \Dom(q_3)\\
                        \Target_{q_3}
                    \end{array}$};
                \end{tikzpicture}
                }
            \end{adjustbox}
            \caption{The $\edHAabb$ }
            \label{fig:edha}
        \end{minipage} 
        }
        \begin{minipage}[b]{0.45\textwidth}
            \centering
            \vspace*{2mm}
            \begin{adjustbox}{max width=1\linewidth}
                \scalebox{1}{
                    \begin{tikzpicture}[mode/.style={circle,draw=black,thick,
                        inner sep=0pt,minimum size=5mm},
                        target/.style={circle,draw=black,double,thick,
                        inner sep=0pt,minimum size=5mm},]
                        \node (q12) at (0.8,0) [mode] {\tiny{$q_{11}$}};
                        \node (q13) at (1.8,0.7) [mode] {\tiny{$q_{12}$}};
                        \node (q14) at (0.4,0.5) [mode] {\tiny{$q_{13}$}};

                        \node (q30) at (1.5,2.8) [target] {\tiny{$q_{30}$}};
                        \node (q31) at (2.3,2.5) [mode] {\tiny{$q_{31}$}};

                        \node (q2) at (3.5,0) [mode] {\tiny{$q_{2}$}};
                        \draw [->,thick] (q12) to (q2);
                        \draw [->,thick] (q13) to[bend left = 30] (q30);
                        \draw [->,thick] (q13) to[bend right = 30] (q31);
                        \draw [->,thick] (q14) to (q2);
                        \draw [->,thick] (q14) to[bend left = 30] (q30);
                        \draw [->,thick] (q14) to[bend left = 30] (q31);
                        \draw [->,thick] (q31) to[bend left = 10] (q14);
                        \draw [->,thick] (q31) to [bend right = 10] (q13);
                        \draw [->,thick] (q31) to (q12);


                        \draw [->,thick] (-0.2,0.3) to (q14);
                        \draw [->,thick] (0.2,-0.2) to (q12);
                        \draw [->,thick] (1.2,0.5) to (q13);
                        %

                    \end{tikzpicture}
                }
            \end{adjustbox}
            \caption{The resulting discrete directed graph}
            \label{fig:nfa}
        \end{minipage}
    \end{figure}

\section{Reset Controller Synthesis}\label{sec:discrete}
In this section, we focus on the third step of our approach, which involves synthesizing a reset controller for a given $\edHAabb$ $\mathcal{H}$
utilizing the discrete directed graph obtained from $\mathcal{H}$ in the second step.

\subsection{Pruning  the Resulting Discrete Directed Graph}\label{subsec:modif}

In this subsection, we will present a method to prune some edges that may violate the reach-avoid property from the discrete directed graph derived in the second step by redefining their reset maps.

\begin{small}
    \begin{algorithm}[h]
        \caption{$\mathtt{PruningNonTargetSink}(\DG)$}
        \label{alg:NFADFS}
        \begin{algorithmic}[1]
            \REQUIRE  $\DG=\{V, A, V_0, V_T\}$ 
            \ENSURE  $\DG^\prime=\{V^\prime, A^\prime, V_0^\prime, V_T \}$ without non-target sinks
            \STATE $V^\prime \gets V$, $A^\prime \gets A$, $V_0^\prime \gets V_0$ 
            \STATE $u\gets$ the number of modes in $V$ \label{alg3:line1}
            \STATE construct the adjacent matrix $M=(m_{ij})_{u\times u}$ for $\DG$ \label{alg3:line2}
            \REPEAT \label{alg3:line3}
            \STATE $S\gets\left\{q_s\notin V_T \left|\,\sum_{i=1}^{u} m_{si}=0\right.\right\}$, flag $\gets 0$ \label{alg3:line4}
            \FOR {$q_s\in S$} \label{alg3:line5}
            \FOR {$i=1$ \textbf{to} $u$} \label{alg3:line6}
            \STATE
            \textbf{if} {$(q_i,q_s)\in A$}
            \textbf{then}
            $A' \gets A \setminus
                \{(q_i,q_s)\}$,
            \quad \quad \STATE $m_{is}\gets 0$,flag $\gets 1$ \label{alg3:line7}
            \ENDFOR \label{alg3:line8}
            \STATE $V'\gets V^\prime \setminus
                \{q_u\}, V_0'\gets
                V_0^\prime  \setminus
                \{q_u\}$
            \ENDFOR \label{alg3:line9}
            \UNTIL {flag=0} \label{alg3:line10}
            \RETURN $\DG^\prime=\{V^\prime, A^\prime, V_0', V_T\}$ \label{alg3:line11}
        \end{algorithmic}
    \end{algorithm}
\end{small}

As shown in Fig.\nobreakspace \ref {fig:nfa} in Example\nobreakspace \ref {exp:trans}, there are two types of scenarios that may violate the reach-avoid property:
\begin{itemize}
    \item[1.] The first case is ``non-target sink''. A sink refers to a vertex with no outgoing edges. This occurs when a path terminates at a sink that does not belong to $V_T$. An example is the path $\langle q_{11}, q_{2}\rangle$ in Fig.\nobreakspace \ref {fig:nfa}, which terminates at $q_{2}$ with $q_{2} \not \in V_T$. To address this issue, our pruning algorithm will iteratively remove all non-target sinks and their incoming arcs until none non-target sinks are left.(Alg.\nobreakspace \ref {alg:NFADFS}). For instance, Alg.\nobreakspace \ref {alg:NFADFS} will prune $(q_{11},q_2)$, $(q_{13},q_2)$, $(q_{31},q_{11})$ and $(q_{31},q_{13})$ in Fig.\nobreakspace \ref {fig:nfa}.
    \item[2.] The second type is referred to as  ``simple loop''. This occurs when the discrete directed graph contains a simple loop, in which at most one vertex can appear twice, such as $\langle q_{12},q_{31}, q_{12} \rangle$ in Fig.\nobreakspace \ref {fig:nfa}. Clearly, such path could be executed infinitely. So it is impossible to reach any target in case no target vertex is contained in the loop.
        To address this issue, we propose another algorithm (Alg.\nobreakspace \ref {alg:NFAResetSyn}) to break simple loop by removing some edges on the path backwards. For instance, in Fig.\nobreakspace \ref {fig:nfa}, we prune  $(q_{12},q_{31})$ and $(q_{31},q_{12})$ in order to break the simple loop $\langle q_{12},q_{31}, q_{12} \rangle$.
\end{itemize}



Alg.\nobreakspace \ref {alg:NFADFS}  
iteratively removes non-target sinks and their incoming arcs. With the aid of adjacent matrix given by line \ref{alg3:line3}, such vertex $q_s\notin V_T$ can be detected if $\sum_{i=1}^{u} m_{si}=0$ (line \ref{alg3:line5}). Next, Alg.\nobreakspace \ref {alg:NFADFS} removes all transitions to it, and in case new vertices without incoming transitions are created in the removal process, we conduct this part iteratively until no non-target sinks left.
Alg.\nobreakspace \ref {alg:NFAResetSyn} is to address the two previously mentioned cases by blocking simple loops that do not reach the target set and pruning non-target sinks.
To begin with, the algorithm iteratively identifies simple loops within $\DG^\prime$, and removes a selected edge from each loop (line \ref{alg4:prun_edge}). This process continues until there are no more simple loops present in $\DG^\prime$. Subsequently, Alg.\nobreakspace \ref {alg:NFAResetSyn} is applied to prune all non-target sinks within the resulting modified discrete directed graph (line \ref{alg4:line6}). Finally, the algorithm returns the modified discrete directed graph as the output (line \ref{alg4:line7}).
\begin{small}
    \begin{algorithm}[h]
        \caption{$\mathtt{PruningGraph}(\DG)$}
        \label{alg:NFAResetSyn}
        \begin{algorithmic}[1]
            \REQUIRE  $\DG=\{V, A, V_0, V_T\}$
            \ENSURE  $\DG^\prime=\{V^\prime, A^\prime, V_0^\prime, V_T^\prime\}$ satisfying the reach-avoid constraints
            \STATE $V^\prime\gets V$, $A^\prime\gets A$, $V_0^\prime\gets V_0$, $V_T^\prime\gets V_T$
            \label{alg4:line1}
            \REPEAT
            \STATE Find a simple loop $c$ in $\DG^\prime$
            \STATE $A^\prime\gets A^\prime\setminus e$, where $e=(p,q)\in A^\prime$ is an edge in $c$\label{alg4:prun_edge}
            \UNTIL{There is no simple loop in $\DG^\prime$}
            \STATE $\DG^\prime\gets\mathtt{PruningNonTargetSink}(\DG^\prime)$ \label{alg4:line6}
            \RETURN $\DG^\prime=\{V^\prime, A^\prime, V_0^\prime, V_T^\prime\}$ \label{alg4:line7}
        \end{algorithmic}
    \end{algorithm}
\end{small}

The following theorem gives the correctness of Algorithm \ref{alg:NFAResetSyn}.
    \begin{thm}\label{thm:alg1-correct}
        Algorithm \ref{alg:NFAResetSyn} is correct, meaning that:
        \begin{itemize}
            \item \textbf{Termination:} Algorithm \ref{alg:NFAResetSyn} always terminates.
            \item \textbf{Soundness:} The algorithm outputs a pruned DDG in which all traces terminate at target vertices.
            \item \textbf{Completeness:} If the original DDG contains traces that terminate at target vertices, the algorithm always produces a non-empty DDG.
        \end{itemize}
    \end{thm}
    \begin{pf}
        \textbf{Termination.} Since the number of edges and vertices decreases by at least one in each iteration, the pruning process will eventually terminate. Therefore, Algorithm \ref{alg:NFAResetSyn} terminates for any DDG.

        \textbf{Soundness.} After removing all simple loops in Algorithm \ref{alg:NFAResetSyn} (lines 2-4), any remaining path is finite. Suppose $v_0v_1\dots v_t$ is a finite path in $\DG^\prime$ such that $v_0 \in V_0$ and $v_t$ has no successor. We need to show if $v_t \in V_t$ then the finite path reaches a target vertex. Line 6 in Algorithm \ref{alg:NFAResetSyn} ensures that $v_t$ is a target vertex because Algorithm \ref{alg:NFADFS} iteratively removes all non-target vertices that do not have successors (cf. lines 4-12).

        \textbf{Completeness.} By the definition of DDG from a dHA, target vertices have no outgoing arcs. Thus, target vertices will not be included in any simple loop, so line 4 of Algorithm \ref{alg:NFAResetSyn} will not prune traces that terminate at target vertices. Additionally, traces that terminate at target vertices are not non-target sinks, ensuring that these traces are always retained.
        \qed
    \end{pf}

\subsection{Reset Controller Synthesis from Discrete Directed Graph}\label{subsec:transback}

In this subsection, our focus is on synthesizing a reset controller based on the resulting discrete directed graph $\DG'$ obtained from applying Alg.\nobreakspace \ref{alg:NFADFS} and Alg.\nobreakspace \ref{alg:NFAResetSyn}. The goal is to generate a reset map associated with each edge in the original $\edHAabb$ $\edHAsyb$. This is achieved by merging the reset maps derived from the mode partitioned $\edHAabb$ $\edHAsyb^m$ using Alg.\nobreakspace \ref{alg:mode-seg}, and then refining the initial set and restricting the reset map according to the structure of $\DG'$.

Let's continue the running example to illustrate the basic idea in the following. 

\begin{exmp}
    By applying Alg.\nobreakspace \ref {alg:NFAResetSyn} to Example\nobreakspace \ref {exp:trans}, the
    resulting discrete directed graph $\DG'$ is presented in Fig.\nobreakspace \ref {fig:modi-NFA}.
    In the following, we will use $\Reset^m(\cdot)$ and $\Init^m(\cdot)$ to represent the resulting reset map and initial set  from the running example by applying  Alg.\nobreakspace \ref {alg:mode-seg}.
    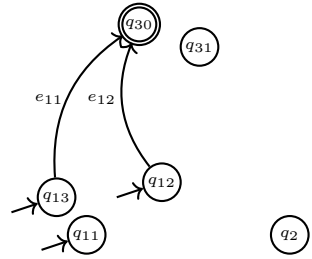
\begin{figure}[h]
        \centering
        \begin{adjustbox}{max width=1\linewidth}
            \scalebox{1}{
                \begin{tikzpicture}[mode/.style={circle,draw=black,thick,
                                inner sep=0pt,minimum size=5mm},
                        target/.style={circle,draw=black,double,thick,
                                inner sep=0pt,minimum size=5mm},]
                    \node (q12) at (0.8,0) [mode] {\tiny{$q_{11}$}};
                    \node (q13) at (1.8,0.7) [mode] {\tiny{$q_{12}$}};
                    \node (q14) at (0.4,0.5) [mode] {\tiny{$q_{13}$}};

                    \node (q30) at (1.5,2.8) [target] {\tiny{$q_{30}$}};
                    \node (q31) at (2.3,2.5) [mode] {\tiny{$q_{31}$}};

                    \node (q2) at (3.5,0) [mode] {\tiny{$q_{2}$}};
                    \draw [->,thick] (q13) to[bend left = 30] (q30);
                    \draw [->,thick] (q14) to[bend left = 30] (q30);

                    \node [below of=q30,xshift=-1.2cm,yshift=0cm] {\tiny{$e_{11}$}};
                    \node [below of=q30,xshift=-0.5cm,yshift=0cm] {\tiny{$e_{12}$}};

                    \draw [->,thick] (-0.2,0.3) to (q14);
                    \draw [->,thick] (0.2,-0.2) to (q12);
                    \draw [->,thick] (1.2,0.5) to (q13);
                \end{tikzpicture}
            }
        \end{adjustbox}
        \caption{The discrete directed graph of Example\nobreakspace \ref {exp:trans} after applying Alg.\nobreakspace \ref {alg:NFAResetSyn}}
        \label{fig:modi-NFA}
    \end{figure}      %
    \begin{itemize}
        \item $\Reset^r(e_2,\Guard(e_2)\setminus\Guard(e_1))$\footnote{the $e_1$, $e_2$ here is the edge of $\edHAabb$ labeled in Fig.\nobreakspace \ref {fig:hldo}}$=\Reset^m(e_{12},\Guard^m(e_{12}))$, corresponding to  $e_{12}$ in Fig.\nobreakspace \ref {fig:modi-NFA}; 
        \item $\Reset^r(e_2,\Guard(e_2)\cap\Guard(e_1))=\Reset^m(e_{11},\Guard^m(e_{11}))$, corresponding to $e_{32}$ in Fig.\nobreakspace \ref {fig:modi-NFA};
        \item $\Init^r(q_1)=\Init^m(q_{13})\cup\Init^m(q_{14})$, corresponding to  $e_{11}$ and $e_{12}$ in Fig.\nobreakspace \ref {fig:modi-NFA};
        \item $\Reset^r$ will associate null map  with other edges and
              $\Init^r$ assigns initial sets in other mode to be empty.
    \end{itemize}

    Thus, with the re-defined reset map and initial set, all executions of $\edHAabb$ in Example\nobreakspace \ref {exp:running} will reach to the target set within $q_2$, while ensuring safety before the reaching.
\end{exmp}

We implement all above ideas
together in Alg.\nobreakspace \ref {alg:reset-syn}. Line \ref{alg5:deterministic} applies
Alg.\nobreakspace \ref {alg:mode-seg} to
obtain a deterministic  $\edHAabb$ $\edHAsyb^m$. Line~\ref{alg5:refine} obtains a discrete directed graph $\DG$ from $\edHAsyb^m$. Then,
line~\ref{alg5:pruningSL} prunes all simple loops and non-target sinks that cannot reach the target set using Alg.\nobreakspace \ref {alg:NFAResetSyn}.
Afterwards, Alg.\nobreakspace \ref {alg:reset-syn} tries
to redefine a reset map and refine the initial set by traversing  all modes of $\edHAsyb$ (line~\ref{alg5:reset-start} to line~\ref{alg5:reset-end}). In each iteration, the initial set of the selected mode $q_i$ is set to be empty first (line~\ref{alg5:init_init});  for each sub-mode $q_{ij}$ of $q_i$,
$P$ collects all submodes $q_{kl}$ from $\DG'$, where there is an edge
between $q_i$ and $q_k$ in $\edHAsyb$, and then the reset map
associated with $(q_i,q_k)$ is set to be the union of
all reset maps associated with $(p_{ij},p)$ in $\DG'$, where $p\in P$;
correspondingly, the initial set of $q_i$ is set to the union of the initial set of $q_{ij}$, where $q_{ij}$ in $\DG'$ is a submode of $p_i$.

\begin{small}
    \begin{algorithm}[h]
        \caption{Reset Control Synthesis for $\edHAabb$}
        \label{alg:reset-syn}
        \begin{algorithmic}[1]
            \REQUIRE  A  $\edHAabb$ $\edHA$, safe set $\Safe\subseteq\DisState\times\ContiState$, and  target set $\Target\subseteq\DisState\times\ContiState$
            \ENSURE  $\edHAsyb^r~=~(\DisState,\ContiState,\Dom,\ContiDyn,\Init^r,\DisJump,\Guard,\STime,\Reset^r)$ satisfying the safety together with the liveness constraints.
            \STATE $\edHAsyb^m, \,\Target^m,\,\Safe^m
                \gets \ModeSeg{\edHAsyb}{\Safe}{\Target}$
            \label{alg5:deterministic}
            \STATE $\DG \gets \{V,A,V_0,V_T\}$\label{alg5:refine}
            \STATE $\DG'=\{V',A',V_0',V_T'\}\gets \mathtt{PruningGraph}(\DG)$  \label{alg5:pruningSL}
            \FOR{$q_i\in\DisState$ of $\edHAsyb$ }\label{alg5:reset-start}
            \STATE $\Init^r(q_i)\gets \emptyset$ \label{alg5:init_init}
            \FOR {each sub-mode $q_{ij}$ of $q_i$}\label{alg5:guard-start}
            \STATE $\textit{Post}(q_{ij})\gets\{q_k\in\DisState\mid (q_{ij},q_{kl})\in A'$ with $q_{kl} \in V'$ being a sub-mode of $q_k\}$\\
            \FOR{$q_k\in \textit{Post}(q_{ij})$}\label{alg5:dif_start}
            \STATE $P\gets \{q_{kl} \mid q_{kl}
                \text{ is a sub-mode of }q_k \text{ and }$\\$ (q_{ij},p_{kl})\in A'\}$\\
                /*~\textit{$P$ contains any sub-mode of $q_k$  which has an edge from $q_{ij}$ and is not pruned in Alg.\nobreakspace \ref {alg:NFAResetSyn}}~*/
                \STATE $\Reset^r((q_i,q_k),\Guard^m(q_{ij},q_{kl}))$\\
                \qquad\qquad\qquad$\gets\bigcup_{p\in P}\Reset^m((q_{ij},p),\Guard(q_{ij},p))$\label{alg5:reset}
                \ENDFOR\label{alg5:dif_end}
                \IF{$Post(q_{ij})\neq \emptyset$}\label{alg5:init_start}
                \STATE $\Init^r(q_i)\gets\Init^r(q_i)\cup\Init^m(q_{ij})$
                \ENDIF\label{alg5:init_end}
                \ENDFOR\label{alg5:guard-end}
                \ENDFOR\label{alg5:reset-end}
                \IF{$\Init^r\neq\emptyset$}
                \RETURN $\edHAsyb^r~=~(\DisState,\ContiState,\Dom,\ContiDyn,\Init^r,\DisJump,\Guard,\STime,\Reset^r)$
            \STATE \textbf{else} \textbf{return} "Synthesis fail"
            \ENDIF
        \end{algorithmic}
    \end{algorithm}
\end{small}
\quad

\begin{thm}[Correctness]\label{thm:correct-complete}
    The reset controller synthesis problem for dHA is solvable if and only if the resulting $\Init^r$ obtained from the synthesis procedure is non-empty.
    \begin{itemize}
        \item \textbf{Soundness}: Any reset controller synthesized by our method effectively solves the problem.
        \item \textbf{Conditional Completeness}: If the reach-avoid set in each mode can be explicitly calculated and a reset controller exists, our method will synthesize it.
    \end{itemize}
\end{thm}

The above theorem gives a theoretical guarantee of our method, which is a straightforward consequence of Theorems \ref{thm:alg1-correct} and \ref{thm:inner-approx-ra}.

\section{Implementation and Experiments}\label{sec:exper}

To further illustrate the efficacy of our approach,
we implement  a prototypical tool using \textsc{Matlab} (2022b) and Python. Alg.~\ref{alg:mode-seg} is implemented in \textsc{Matlab} (2022b), integrated with \textsc{Yalmip} \cite{lfberg2004toolbox} and \textsc{Mosek} \cite{andersen2003implementing} to formulate and solve the underlying SOS constraints. Alg.~\ref{alg:NFADFS} and Alg.~\ref{alg:NFAResetSyn} are implemented in Python, leveraging the NetworkX package to manipulate the resulted discrete directed graph.

We first apply the  tool to a nonlinear system of prey-predator, and subsequently evaluate its performance on a collection of benchmark examples\footnote{Some examples are obtained by duplicating their continuous dynamics with different target sets to create the corresponding delay hybrid systems.} 
in the literature to demonstrate the scalability of our approach. 
All experiments are  conducted on an Apple M2 laptop with 8GB of RAM, operating on macOS Ventura (V13.2). The experimental results are presented in Table\nobreakspace \ref {tab:result}.

    \subsection{A Nonlinear System of Prey-Predator}
    \begin{exmp}\label{exp:Damp-Pla}
        Consider a nonlinear system of pre-predator, where the population dynamics are described by the following DDE:
        \begin{align*}
            \dot x_1(t) & = bx_1(t)b - x_1(t)^2 - 4x_1(t)x_2(t)\\
            \dot x_2(t) & = 4.8x_1(t-\tau)x_2(t-\tau) - x_2(t-\tau)
        \end{align*}
        Here, $x_1(t)$ and $x_2(t)$ represent the populations of prey and predator, respectively; $b$ denotes the rate of increase of the prey population;  $\tau$ accounts for the maturation process of the predator population. We assume $\tau=0.001$ in this example.
        
        The increase rate $b$ of prey may be influenced by the environment and change between different values.  In Fig.\nobreakspace \ref {fig:hldo2}, a $\edHAabb$ is presented to depict such switch, where each mode corresponds to a specific value of $b$. 

        \begin{figure}[h]
            \centering
            \vspace*{-3mm}
            \begin{adjustbox}{max width=0.8\linewidth}
            \scalebox{1}{
                \begin{tikzpicture}[mode/.style={circle,draw=black,thick,
                    inner sep=0pt,minimum size=18mm},
                    target/.style={circle,draw=black,double,thick,
                    inner sep=0pt,minimum size=5mm},]
                    \node (q1) at (-2,0) [mode] {};
                    \node (q2) at (2,0) [mode] {};
                    \node (q3) at (0,3.5) [mode] {};
                    \draw[->,thick] (q1) to[bend right = 15] (q2);
                    \node [below of=q3, xshift=0cm,yshift=-2.8cm] {\scriptsize $\begin{array}{c}
                        e_1,G(e_1)\\
                        R(e_1,(x_1,x_2))
                    \end{array}$};
                    \draw[->,thick] (q2) to[bend right = 15](q3);
                    \node [below of=q3, xshift=2.2cm,yshift=-0.5cm] {\scriptsize $\begin{array}{c}
                        R(e_2,(x_1,x_2))\\
                        e_2,G(e_2)
                    \end{array}$};
                    \draw[->,thick] (q3) to[bend right = 15] (q1);
                    \node [below of=q3, xshift=-2.2cm,yshift=-0.5cm] {\scriptsize $\begin{array}{c}
                        R(e_3,(x_1,x_2))\\
                        e_3,G(e_3)
                    \end{array}$};

                    \node [below of=q1,yshift=1.7cm] {$q_1$};
                    \node [below of=q1,yshift=0.9cm] {\scriptsize$\begin{array}{c}
                        b=2\\
                        Dom(q_1)\\
                        Init(q_1)
                    \end{array}$};

                    \node [below of=q2,yshift=1.5cm] {$q_2$};
                    \node [below of=q2,yshift=0.8cm] {\scriptsize$\begin{array}{c}
                        b=1.2\\
                        Dom(q_2)
                    \end{array}$};

                    \node [below of=q3,yshift=1.5cm] {$q_3$};
                    \node [below of=q3,yshift=0.8cm] {\scriptsize$\begin{array}{c}
                        b=0.8\\
                        Dom(q_3)
                    \end{array}$};
                    \node at(0,-3) {\scriptsize$\begin{aligned}
                        &Dom(q_1)  = \{(x_1,x_2)\mid (x_1-0.2)^2 + (x_2-0.2)^2\le 0.04\},\\
                        &Init(q_1) = \{(x_1,x_2)\mid (x_1-0.2)^2 + (x_2-0.2)^2\le 0.04\},\\
                        &G(e_1)    = \{(x_1,x_2)\mid (x_1-0.4)^2 + (x_2-0.3)^2\le 0.01\},\\
                        &Dom(q_2)  = \{(x_1,x_2)\mid (x_1-0.5)^2 + (x_2-0.5)^2\le 0.25 \},\\
                        &G(e_2)    = \{(x_1,x_2)\mid (x_1-0.1)^2 + (x_2-0.5)^2\le 0.01\},\\
                        &Dom(q_3)  = \{(x_1,x_2)\mid (x_1-0.3)^2 + (x_2-0.3)^2\ge 0.09\},\\
                        &G(e_3)    = \{(x_1,x_2)\mid (x_1-0.1)^2 + (x_2-0.5)^2\le 0.01\}
                    \end{aligned}$};
                \end{tikzpicture}
                }
            \end{adjustbox}
            \caption{The $\edHAabb$ for the prey-predator system}
            \label{fig:hldo2}
        \end{figure}       %

        The safe set in all modes is equal to its domain, formally, $\Safe_{q_1}=Dom(q_1)$, $\Safe_{q_2}=Dom(q_2)$, $\Safe_{q_3}=Dom(q_3)$; the target set in $q_3$ is defined as $\Target_{q_3}=\{(x_1,x_2) \mid (x_1-0.2)^2 + (x_2-0.2)^2 \le 0.01\}$ and the target set in $q_1$, $q_2$ is empty. 

        Denote the resulted $\edHAabb$ with Alg.\nobreakspace \ref {alg:reset-syn} by $\edHAsyb^r~=~(\DisState,\ContiState,\Dom,\ContiDyn,\allowbreak\Init^r,\allowbreak\DisJump,\allowbreak\Guard,\allowbreak\STime,\Reset^r)$. 
        For simplicity, 
        we use $\RA_{in}(i,j)$ to denote the reach-avoid set approximated by Alg.\nobreakspace \ref {alg:inner-ra}, i.e.,
        \begin{align*}
            &\RA_{in}(1,1) \DefSym \InnerRA{\f_1}{Dom(q_1)\cap S_{q_1}}{G(e_1)\setminus Dom(q_1)},\\
            &\RA_{in}(2,1) \DefSym \InnerRA{\f_2}{Dom(q_2)\cap S_{q_2}}{G(e_2)\setminus Dom(q_2)},\\
            &\RA_{in}(3,1) \DefSym \InnerRA{\f_3}{Dom(q_3)\cap S_{q_3}}{G(e_3)\setminus Dom(q_3)},\\
            &\RA_{in}(3,0) \DefSym \InnerRA{\f_3}{Dom(q_3)\cap S_{q_3}}{T_{q_3}}
        \end{align*}
        Then, the reset controller we synthesized is given by 
        \begin{align*}
            &\Reset^r(e_1,\Guard(e_1)) = \RA_{in}(2,1),\\
            &\Reset^r(e_2,\Guard(e_2)) = \RA_{in}(3,0),\\
            &\Reset^r(e_3,\Guard(e_3)) = \emptyset,\\
            &\Init^r(q_1) = \Init(q_1) \cap \RA_{in}(1,1)
        \end{align*}

        \begin{figure}[h]
            \centering
                \begin{adjustbox}{max width=0.9\linewidth}
                    \scalebox{1}{
                        \begin{tikzpicture}
                            \draw (0, 0) node[inner sep=0] {\includegraphics[width=1\linewidth,trim=8cm 0cm 2cm 0cm, clip]{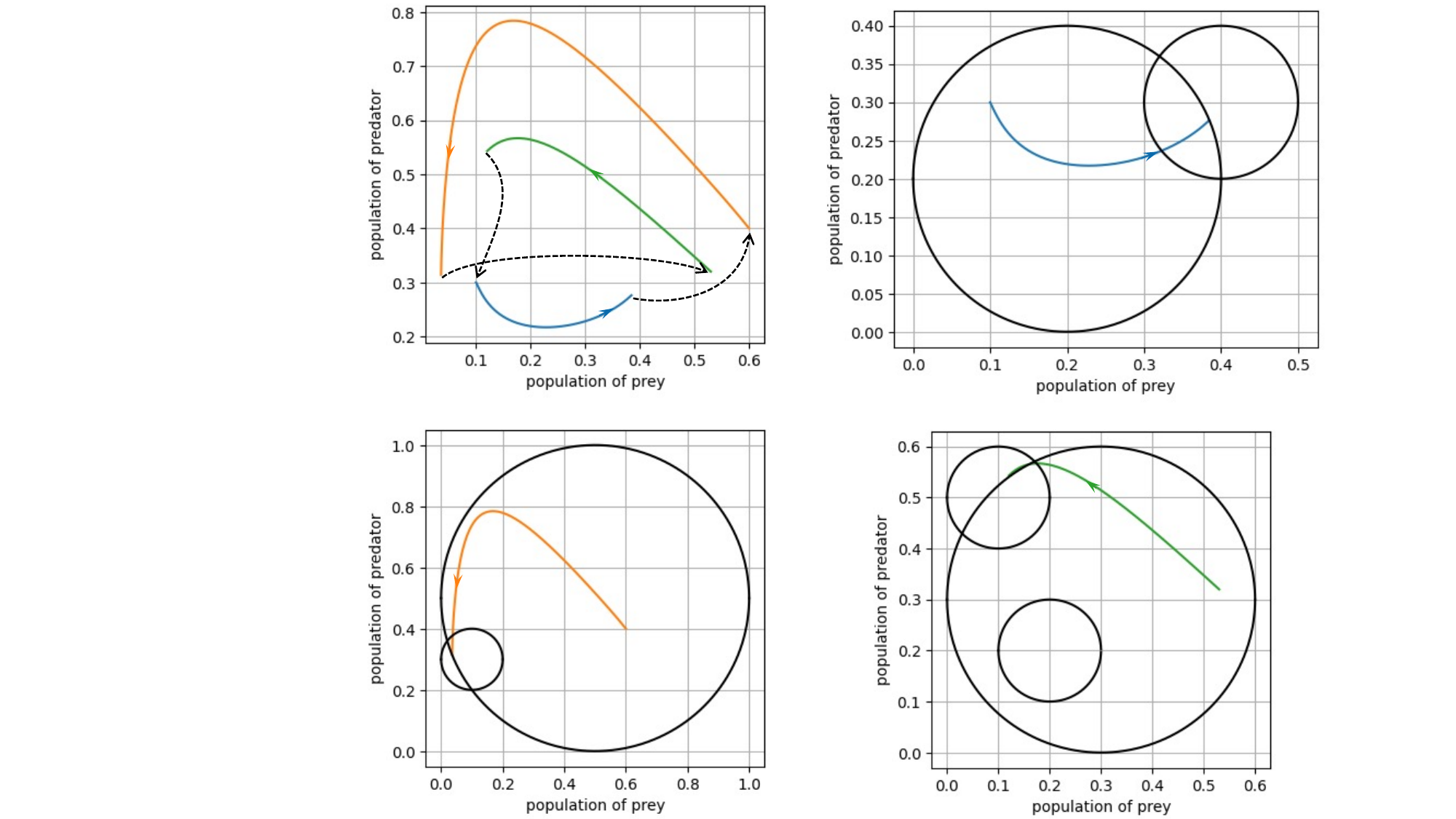}};
                            \node at(-2.3,0) {\scriptsize (a)};
                            \node at(2,0) {\scriptsize (b)};
                            \node at(-2.3,-3.5) {\scriptsize (c)};
                            \node at(2,-3.5) {\scriptsize (d)};

                            \node at(-1.2,0.75) {\tiny{$R(\!e_1\!,\!G(\!e_1\!)\!)$}};
                            \node at(-2.7,1.8) {\tiny{$R(\!e_2\!,\!G(\!e_2\!)\!)$}};
                            \node at(-2.3,1.1) {\tiny{$R(\!e_3\!,\!G(\!e_3\!)\!)$}};
                            \node at(-2.5,0.8) {\tiny{$q_1$}};
                            \node at(-2,3) {\tiny{$q_2$}};
                            \node at(-1.8,1.9) {\tiny{$q_3$}};

                            \node at(2,1) {\tiny{$D\!o\!m(q_1)$}};
                            \node at(3.2,2.8) {\tiny{$G(e_1)$}};
                            \node at(-2.3,-2.5) {\tiny{$D\!o\!m(q_2)$}};
                            \node at(-2.7,-1.8) {\tiny{$G(e_2)$}};
                            \node at(2.6,-2) {\tiny{$D\!o\!m(q_3)$}};
                            \node at(1.15,-0.7) {\tiny{$G(e_3)$}};
                            \node at(1.6,-2) {\tiny{$T_{q_3}$}};
                        \end{tikzpicture}
                    }
                \end{adjustbox}
                \caption{Simulation of the pre-predator system before reset controller synthesis: Figure (a) shows the simulation of the pre-predator system before reset controller synthesis, where an execution begins from the initial set $Init(q_1)$. The system runs indefinitely within the safe region but does not enter the target set $T_{q_3}$. The dashed line in Figure (a) represents the reset map used in the simulation. For further clarity, Figures (b), (c), and (d) illustrate the execution in modes $q_1$, $q_2$, and $q_3$, respectively. 
                }
                \label{fig:simul_before}
        \end{figure} 

        \begin{figure}[h]
            \centering
                \begin{adjustbox}{max width=0.9\linewidth}
                    \scalebox{1}{
                        \begin{tikzpicture}
                            \draw (0, 0) node[inner sep=0] {\includegraphics[width=1\linewidth,trim=8cm 0cm 2cm 0cm, clip]{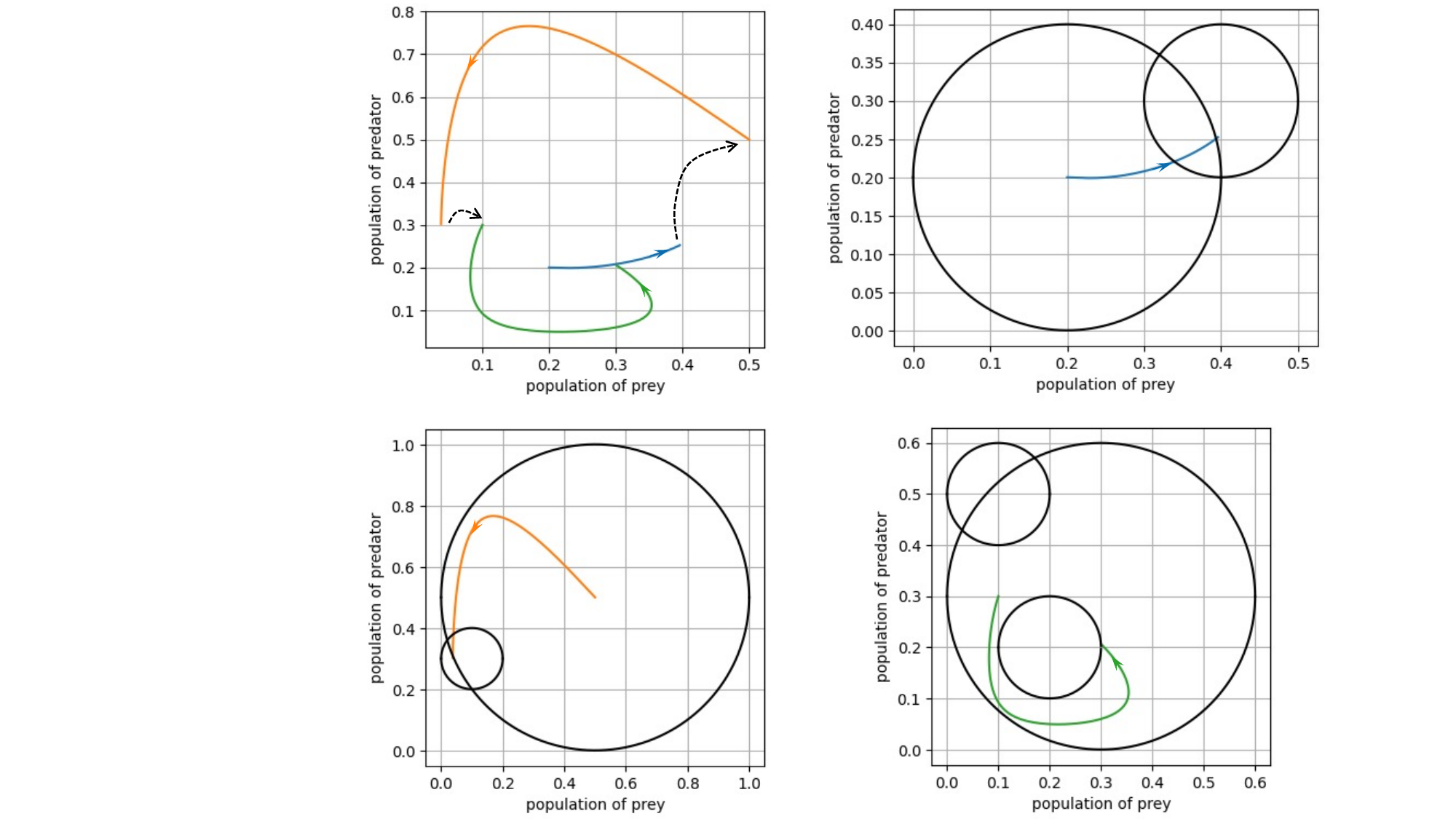}};
                            \node at(-2.3,0) {\scriptsize (a)};
                            \node at(2,0) {\scriptsize (b)};
                            \node at(-2.3,-3.5) {\scriptsize (c)};
                            \node at(2,-3.5) {\scriptsize (d)};

                            \node at(-1.9,2.3) {\tiny{$R^r\!(\!e_1\!,\!G(\!e_1\!)\!)$}};
                            \node at(-2.7,1.8) {\tiny{$R^r\!(\!e_2\!,\!G(\!e_2\!)\!)$}};
                            \node at(-1.8,1.3) {\tiny{$q_1$}};
                            \node at(-2,3) {\tiny{$q_2$}};
                            \node at(-3,0.85) {\tiny{$q_3$}};

                            \node at(2,1) {\tiny{$D\!o\!m(q_1)$}};
                            \node at(3.2,2.8) {\tiny{$G(e_1)$}};
                            \node at(-2.3,-2.5) {\tiny{$D\!o\!m(q_2)$}};
                            \node at(-2.7,-1.8) {\tiny{$G(e_2)$}};
                            \node at(2.8,-1.5) {\tiny{$D\!o\!m(q_3)$}};
                            \node at(1.15,-0.7) {\tiny{$G(e_3)$}};
                            \node at(1.6,-2) {\tiny{$T_{q_3}$}};
                        \end{tikzpicture}
                    }
                \end{adjustbox}
                \caption{Simulation of the pre-predator system after reset controller synthesis: Figure (a) presents an execution starting from the initial set $Init^r(q_1)$ and terminating at the target set $T_{q_3}$. The dashed line in Figure (a) represents the reset map used in the simulation. Figures (b), (c), and (d) provide an illustration of the execution in mode $q_1$, $q_2$, and $q_3$ respectively. 
                }
                \label{fig:simul}
        \end{figure}   
    
        Simulations of the pre-predator system before and after reset controller synthesis is present in Fig.\nobreakspace \ref{fig:simul_before} and Fig.\nobreakspace \ref {fig:simul}, respectively. 
        In Fig.\nobreakspace \ref{fig:simul_before}, it can be observed that the pre-predator system without the synthesized reset controller may infinitely operate within the safe region never entering the target set. In fact, during the progression of the reset controller synthesis, edge $e_3$ has been disabled  from the $\edHAabb$ since all  execution paths through $q_3$ are forced to enter 
        the target set in order to avoid infinitely many iteration among $q_3 \rightarrow q_1 \rightarrow q_2 \rightarrow q_3$. With the synthesized 
        reset map,  all continuous evolutions in $q_3$ will lead  to the target set $\Target_{q_3}$ eventually, meanwhile all execution paths  in each mode stay in the safe set. 
    \end{exmp}

    \subsection{More Case Studies}
    In this subsection, we aim to demonstrate the scalability of our method by applying it to a range of large-scale benchmarks from the literature. The results obtained with our tool on these  benchmark examples  are presented in Table\nobreakspace \ref {tab:result}.
    \begin{table*}[t]
        \begin{center}
        \caption{Experimental results for reset controller synthesis}
        \label{tab:result}
            \begin{tabular}{l  ccccc c ccc c ccc} 
                \toprule
                \multirow{2}{*}{Benchmark} & \multirow{2}{*}{dim} & ~ &\multicolumn{3}{c}{original model}& ~ &\multicolumn{3}{c}{synthesis time}& ~ & \multicolumn{3}{c}{result model}\\
                \cmidrule{4-6} \cmidrule{8-10} \cmidrule{12-14} 
                  &  & ~ & modes & ~ & edges & ~ & time 1 & ~ & time 2 & ~ & modes & ~ & edges \\
                \midrule
                \textsf{Oscillator}\cite{xue2021reach} &  2 & ~ & 3 & ~ & 4 & ~ & 10.34s & ~ & 105.16ms & ~ & 3 & ~ & 3 \\
                \textsf{Low-Pass Filter}\cite{althoff2016implementation} &  2 & ~ & 9 & ~ & 14 & ~ & 27.45s & ~ & 127.20ms & ~ & 8 & ~ & 13 \\
                \textsf{Prey-Predator}\cite{bai2021switching} &  2 & ~ & 12 & ~ & 28 & ~ & 79.01 & ~ & 138.58ms & ~ & 12 & ~ & 27 \\
                \textsf{PD Controller}\cite{goubault2018inner} & 2 & ~ & 16 & ~ & 31 & ~ & 111.44s & ~ & 120.78ms & ~ & 15 & ~ & 20 \\
                \textsf{CWH Equation}\cite{homchaudhuri2016computing} &  4 & ~ & 5 & ~ & 8 & ~ & 17m33s & ~ & 112.76ms & ~ & 5 & ~ & 7\\
                \bottomrule
            \end{tabular}
        \end{center}
        \vspace*{-\baselineskip}
        \vspace*{4mm}
        
        \scriptsize{ dim: the dimension of instant states; modes: the number of modes; edges: the number of discrete transitions; 
          original model: the $\edHAabb$ before synthesizing (i.e., $\edHAsyb$); resulted model: the $\edHAabb$ after synthesizing (i.e., $\edHAsyb^r$);
          time 1: the time consumed  by mode partition; 
          time 2: the time consumed by synthesizing reset controller } 
    \end{table*} 
    
    \textsf{Oscillator}, is a hybrid damped oscillator similar to the one in Example\nobreakspace \ref {exp:running}, with more discrete transitions between modes.  \textsf{Low-Pass Filter} is taken from  \cite{althoff2016implementation}, which  consists of 9 low-pass filters and 14 discrete transitions between different modes.  \textsf{Prey-Predator} is a linearized verion of 
    the prey-predator system in Example~\ref{exp:Damp-Pla} with  a larger number of modes and edges. \textsf{PD Controller} is an adaptation of the PD controller presented in \cite{goubault2018inner} and showcases a PD controller designed for a self-driving car that can switch between different modes. Lastly, \textsf{CWH Equation} is a delay hybrid system utilized to model the relative orbital motion of a chase spacecraft in relation to a reference spacecraft. It can switch between different chase strategies.

    The table demonstrates the effectiveness of our method in synthesizing reset controllers for various benchmarks from the literature. The time required for synthesizing  is mainly determined by the mode partition step (time 1), which involves in solving a series of SDP problems for each mode with respect to each guard condition. Therefore, more the synthesis time for examples with higher dimension, which also depend  on the efficiency of the SDP solver used. 
    The time required to synthesize a reset controller in the first step  (time 2) is not sensitive to the size of problems, as the resulted discrete directed diagrams in Table\nobreakspace \ref {tab:result} are relatively small in compared with those usually tackled by NetworkX.
    The advantage of our method  for reach-avoid analysis is its ability to handle examples with larger time delay efficiently. Therefore, we do not include the comparison with the method of \cite{xue2021reach} in the table.

\section{Conclusion}\label{sec:conclu}
In this paper, we provide a constructive method for synthesizing reset controllers for delay hybrid systems, subject to reach-avoid properties. Our approach employs a novel method for inner-approximating reach-avoid sets for polynomial delay differential equations. We then propose a sound and complete method \footnote{Under the assumption that RABFals synthesized with SDP-based approaches are correct.} for synthesizing the reset controller. Through experiments on a set of relevant examples from the literature, we demonstrate the effectiveness of our approach.

In future work, we aim to extend our approach to more general delay hybrid systems (HSs) with more complicated vector fields in each mode. Additionally, we plan to investigate the potential of a correct-by-construction framework for HSs by integrating feedback controller synthesis, switching logic controller synthesis, and reset controller synthesis into a unified approach.

\bibliographystyle{plain}        
\bibliography{references}           



\end{document}